\documentclass[reprint,
 superscriptaddress,
nofootinbib,
 amsmath,amssymb,
 aps,
pra,
]{revtex4-2}

\usepackage[utf8]{inputenc} 
\usepackage[T1]{fontenc}    
\usepackage{hyperref}       
\usepackage{url}            
\usepackage{booktabs}       
\usepackage{amsfonts}       
\usepackage{nicefrac}       
\usepackage{microtype}      
\usepackage{graphicx}
\usepackage{xcolor, etoolbox}
\usepackage{amsmath,amsfonts,amsthm} 
\usepackage{mathtools}
\usepackage{ragged2e}
\usepackage{braket}      
\usepackage[english]{babel} 
\usepackage{enumitem}
\usepackage{xspace}
\usepackage{dsfont}
\usepackage{bm}
\usepackage{bbm}
\usepackage{dcolumn}
\usepackage{mathrsfs}
\usepackage{todonotes}
\usepackage{float}          

\newcommand{\Tr}{\textbf{Tr}}

\hypersetup{
    colorlinks,
    linkcolor=blue,
    citecolor=blue,
    urlcolor=blue,
    breaklinks=true 
}

\DeclarePairedDelimiter\abs{\lvert}{\rvert}

\begin{document}

\title{Entropy estimation in a spin-1 Bose-Einstein condensate}

\author{Yannick Deller}
    \affiliation{Kirchhoff-Institut f\"{u}r Physik, Universit\"{a}t Heidelberg, Im Neuenheimer Feld 227, 69120 Heidelberg, Germany}

\author{Martin G\"{a}rttner}
    \affiliation{Institut für Festkörpertheorie und Optik, Friedrich-Schiller-Universität Jena, Max-Wien-Platz 1, 07743 Jena, Germany}

\author{Tobias Haas}
    \email{tobias.haas@ulb.be}
    \affiliation{Centre for Quantum Information and Communication, École polytechnique de Bruxelles, CP 165, Université libre de Bruxelles, 1050 Brussels, Belgium}

\author{Markus K. Oberthaler}
    \affiliation{Kirchhoff-Institut f\"{u}r Physik, Universit\"{a}t Heidelberg, Im Neuenheimer Feld 227, 69120 Heidelberg, Germany}

\author{Moritz Reh}
    \affiliation{Kirchhoff-Institut f\"{u}r Physik, Universit\"{a}t Heidelberg, Im Neuenheimer Feld 227, 69120 Heidelberg, Germany}
    \affiliation{Institut für Festkörpertheorie und Optik, Friedrich-Schiller-Universität Jena, Max-Wien-Platz 1, 07743 Jena, Germany}

\author{Helmut Strobel}
    \affiliation{Kirchhoff-Institut f\"{u}r Physik, Universit\"{a}t Heidelberg, Im Neuenheimer Feld 227, 69120 Heidelberg, Germany}    

\begin{abstract}
We investigate the information extractable from measurement distributions of two non-commuting spin observables in a multi-well spin-1 Bose-Einstein condensate. We provide a variety of analytic and numerical evidence that suitably chosen classical entropies and classical mutual informations thereof contain the typical feature of quantum entropies known in quantum field theories, that is, the area law, even in the non-Gaussian regime and for a non-zero temperature. Towards a feasible experimental implementation, we estimate entropic quantities from a finite number of samples without any additional assumptions on the underlying quantum state using $k$-nearest neighbor estimators.
\end{abstract}

\maketitle

\section{Introduction}
\label{sec:Introduction}

The scaling of entropic measures associated with the quantum state of a spatial subregion is one of the central characteristics describing how quantum information and entanglement are encoded in spacetime. Originally discovered by Bekenstein in the context of black hole physics \cite{Bekenstein1972,Bekenstein1973} (see also \cite{Hawking1975,Bombelli1986,Srednicki1993,Callan1994}), the entropy of a subregion is, to leading order, proportional to the area of its enclosing surface -- rather than the subregion's volume -- which is conveniently referred to as the \textit{area law}. Over the last two decades, much theoretical evidence has been gathered for its appearance in various contexts, including, for instance, quantum field theory \cite{Calabrese2004,Calabrese2006,Hastings2007,Calabrese2009,Casini2009,Hollands2018}, quantum many-body systems \cite{Amico2008,Peschel2009,Eisert2010}, tensor networks \cite{Schuch2008} and thermalization \cite{Abanin2019}. Also, it has been shown to occur for measures of correlations between the subregion of interest and its complement, e.g., the quantum mutual information \cite{Wolf2008}.

Considerably less evidence is available on the experimental side, which can be traced back to the notoriously difficult task of extracting the local quantum state, i.e., performing full quantum state tomography. So far, the area law has been observed only for a handful of degrees of freedom for which entropic measures can be reduced to direct observables. This includes, for example, a study of a few-site Bose-Hubbard system where the Rényi-2 quantum entropy is read out via two-copy interference \cite{Islam2015,Kaufman2016} and a trapped ion quantum simulator of ten qubits \cite{Brydges2019}, in which the same quantity was extracted using random measurements \cite{Elben2018}. 

The task of accessing the quantum state is even more challenging for continuous quantum many-body systems, which are, in principle, described by quantum fields over continuous positions as well as infinite-dimensional local Hilbert spaces. A recent ultracold atom experiment reported an area law for the quantum mutual information under the assumption that the underlying quantum state is of Gaussian form, in which case measurements of two-point correlation functions suffice to calculate entropic quantities \cite{Tajik2023}. Nevertheless, studies beyond the Gaussian case have remained elusive so far.

What all aforementioned approaches, theoretical as well as experimental, have in common is their reliance on a \textit{quantum} entropy as an indicator for the area law-like behavior of quantum correlations. However, it has been shown recently that the appearance of the area law is by no means restricted to such quantum entropies: when considering phase-space representations and measurement distributions of the quantum state instead, their corresponding \textit{classical} entropies reveal the area law in the \textit{next-to-leading} order terms, i.e. when classical contributions are subtracted properly \cite{Haas2023b}. 

Prominent examples of such distributions are the Wigner $W$-distribution \cite{Wigner1932}, its marginals and the Husimi $Q$-distribution \cite{Husimi1940,Cartwright1976} (see \cite{Lee1995,Schleich2001,Mandel2013} for reviews on phase-space methods). Especially the latter is of particular interest as it is a non-negative and normalized function in phase space, which allows for well-defined entropic descriptions \cite{Wehrl1978,Wehrl1979,Lieb1978,Lieb2014,Haas2023a,Haas2024}. Further, its usefulness for witnessing entanglement has already been demonstrated theoretically in terms of entropic measures \cite{Haas2021b,Haas2022c,Haas2022d} as well as experimentally \cite{Kunkel2019,Kunkel2021}.

Although all phase-space distributions contain the very same information as the density operator, estimating their associated classical differential entropies is a significantly simpler task than reconstructing the full many-body quantum state and computing its quantum entropy. 
The outcome of an experiment can be thought of as a sample drawn from a given distribution (usually the marginals of the Wigner $W$- or the full Husimi $Q$-distribution are considered), see e.g. \cite{Collett1987,Noh1991,Noh1992,Stenholm1992,Leonhardt1993,Mueller2016,Landon2018,Kunkel2019,Kunkel2021}. In principle, one could then attempt to infer the corresponding distribution from these samples, which is equivalent to state tomography when considering a full phase-space distribution, and calculate its entropy thereafter. However, the need for reconstructing the underlying distribution can be bypassed by employing sophisticated methods to estimate the entropy from the sampled data \textit{directly}. In this context, a widely-used approach is the $k$-nearest neighbor ($k$NN) method, which produces an asymptotically unbiased estimate for the entropy without any assumptions on the underlying distribution \cite{Kozachenko1987,Gyrfi1987,Joe1989, Hall1993,Birge1995,Beirlant1997,Singh2003,Kraskov2004,Goria2005,Sricharan2013,Kandasamy2015,Singh2016,Berrett2016,Gao2018,Lu2020}.

In this work, we consider an experimentally friendly scenario in which the build-up of an area law over time is expected: the post-quench dynamics of a multi-well spin-1 Bose-Einstein condensate (BEC). In what follows, we simulate the dynamics using the truncated Wigner approximation (TWA) and employ the $k$NN method to estimate subtracted classical entropies as well as classical mutual informations of phase-space distributions from sampled data. We provide detailed numerical evidence for the appearance of the area law for classical entropies in a variety of scenarios, including, for instance, non-Gaussian states, varying system sizes, and under typical experimental constraints, e.g., thermal fluctuations and finite sample size. Our simulation results are supported by an analytical model covering the early-time dynamics, which allows for a straightforward evaluation of all entropic quantities. With our findings, we pave the ground for an experimental observation of the area law in a continuous quantum many-body system without any prior assumptions on the underlying quantum state.

\textit{The remainder of this paper is organized as follows}. In \autoref{sec:Spin1BEC}, we introduce the model system of our interest -- a spin-1 Bose-Einstein condensate in the multi-well setup -- with a special focus on the Hamiltonian and the spin observables. We proceed with mapping this system to a continuous-variable quantum system described by canonical commutation relations using the so-called undepleted pump approximation in \autoref{sec:ContinousVariables}. Therein, we also put forward phase-space representations of quantum states and show how these are related to the measurement distributions of the spin observables. The two approaches for simulating the dynamics, i.e., the truncated Wigner approximation for the full Hamiltonian and an approximate, analytically solvable model valid for early times, are discussed in \autoref{sec:Dynamics}. For the latter, we employ methods from Gaussian quantum information theory, which we describe in detail. Thereupon, in \autoref{sec:AreaLaw}, we introduce the necessary background on the area law of quantum and classical entropies in phase space. Then, we describe the $k$NN machinery for estimating entropies from sampled data and benchmark the method for several cases relevant to its application to the system of our interest in \autoref{sec:EntropyEstimation}. Finally, we provide our main results in \autoref{sec:Results}. After a comparison of the analytical model with the TWA approach (\autoref{subsec:CorrelationMatrices}), we show plenty of evidence for the emergence of the area law over time for a variety of classical entropies (\autoref{subsec:ObservationAreaLaw}), which is followed by systematic studies of its robustness. More precisely, we consider strongly non-Gaussian distributions (\autoref{subsec:SystematicsDistributions}), a thermal initial state (\autoref{subsec:SystematicsInitialState}), boundary effects (\autoref{subsec:SystematicsPosition}), varying system size (\autoref{subsec:SystematicsSystemSize}), varying sample size (\autoref{subsec:SystematicsSampleSize}) and different types of boundary conditions (\autoref{subsec:SystematicsBoundaryConditions}). A comprehensive discussion of our results and some future prospects are given in \autoref{sec:Discussion}.

\textit{Notation}. We employ natural units $\hbar = k_{\text{B}} = 1$, denote quantum operators by bold letters, e.g.  $\boldsymbol{\phi}$, and classical variables by normal letters, e.g. $\phi$, (similarly for operations on operators / matrices, e.g. $\Tr \{ \boldsymbol{\rho} \}$ /  $\det \{ \gamma \}$) and equip vacuum expressions with a bar, e.g. $\bar{\mathcal{Q}}$. Further, we use upper indices to refer to a specific well, e.g. $\mathcal{Q}^j$ for the $j$-th well, or to a subsystem of a bipartition $AB$, e.g. $\mathcal{Q}^A$, and put lower indices for the hyperfine levels, e.g. $\boldsymbol{a}_{0}$, or relative modes, e.g. $\boldsymbol{a}_{\pm}$.

\section{Multi-well spin-1 BEC}
\label{sec:Spin1BEC}
We begin with a discussion of the multi-well setup of coupled spin-1 Bose-Einstein condensates. Then, we introduce the Hamiltonian governing the corresponding dynamics and the typical observables in such systems.

\subsection{Setup}
\label{subsec:Setup}
We consider an experimental setup consisting of a spin-1 BEC with ferromagnetic spin coupling, based on Lithium-7 \cite{Huh2020} (see also \cite{Kawaguchi2012,Fujiwara2019,Jepsen2020,Kwon2022}). We assume the condensate's geometry to be quasi-one-dimensional, realizable via tight confinement along the radial direction, which can be implemented experimentally by ensuring that the radial trapping frequency is much larger than the longitudinal one, i.e. $\omega_r \gg \omega_l$. In the longitudinal direction, the BEC is further subjected to an optical lattice potential, which divides the overall BEC into $N$ smaller-size BECs, which we refer to as wells. The depth of the optical lattice controls the potential barrier between neighboring wells. When tuned to sufficiently small energies, atoms can tunnel between neighboring wells, which establishes a nearest-neighbor interaction between the local degrees of freedom. This setup constitutes a discretized approximation to a continuous quantum field, see \autoref{subsubsec:FieldTheoryCorrespondence}.

The internal degrees of freedom encoding the spin-1 system are the three Zeeman levels $m_F = 0, \pm 1$ of the $F=1$ hyperfine manifold of the electronic ground state. Initially, we consider all atoms to be prepared in the $m_F = 0$ mode. Thereafter, spin-changing collisions triggered by off-resonant microwave dressing lead to the side modes being dynamically populated \cite{Kawaguchi2012,Hamley2012,Stamper2013,Kim2021}. As this process is symmetric with respect to the side modes, the linear Zeeman shift cancels for all populated states, and hence, only the quadratic Zeeman shift is relevant for the dynamics.

We consider tunneling to only take place in the side modes and not in the initially macroscopically populated zero mode, such that information exchange between wells through tunneling only takes place after the generation of the side-mode population. Experimentally, this can be achieved, for example, by working in a state-dependent lattice, such that the side modes are trapped more weakly than the zero mode. Another possibility is to tune the spin-changing collisions into resonance with excited states of the side modes and work in a deep optical lattice, such that atoms transferred to the side modes acquire enough kinetic energy to tunnel to neighboring sites of the deep optical lattice, whereas the kinetic energy of atoms in the zero mode remains insufficient for tunneling.

For a schematic of our considered setup, see \cite{PRL}.

\subsection{Modes and Hilbert space}
\label{subsec:ModesAndHilbertSpace}

We are interested in the dynamics of the internal degrees of freedom characterized by the three hyperfine levels $m_F \in \{-1, 0, 1 \}$. To each hyperfine level $m_F$ and well $j$, we associate a set of bosonic creation and annihilation operators
\begin{equation}
    \left[ \boldsymbol{a}^j_{m_F}, \boldsymbol{a}^{j' \dagger}_{m'_F} \right] = \delta^{j j'} \, \delta_{m_F m'_F},
    \label{eq:BosonicCommutationRelations}
\end{equation}
where the upper index labels the $N \in \mathbb{N} < \infty$ wells such that $j \in \{1, ..., N \}$. For a single well $j$, the underlying Hilbert space $\mathcal{H}^j$ is a Fock space constructed from multi-particle states
\begin{equation}
    \ket{n^j_1, n^j_0, n^j_{-1}} \equiv \frac{(\boldsymbol{a}_1^{j\dagger})^{n^j_{1}} (\boldsymbol{a}_0^{j\dagger})^{n^j_{0}} (\boldsymbol{a}_{-1}^{j\dagger})^{n^j_{-1}}}{\sqrt{n^j_{1}! \, n^j_{0}! \, n^j_{-1}!}} \ket{0,0,0},
\end{equation}
with the tensor product notation $\ket{., ., .} = \ket{.} \otimes \ket{.} \otimes \ket{.}$ understood and the number of particles $n_{m_F}^j$ in mode $j$ and level $m_F$ being defined as the eigenvalue of the corresponding particle number operator
\begin{equation}
    \boldsymbol{N}^j_{m_F} = \boldsymbol{a}^{j \dagger}_{m_F} \boldsymbol{a}^{j}_{m_F}.
    \label{eq:ParticleNumberOperator}
\end{equation}
Accordingly, the total number of particles in well $j$ is measured by the operator 
\begin{equation}
    \boldsymbol{N}^j = \sum_{m_F = -1}^{1} \boldsymbol{N}^j_{m_F},
\end{equation}
which has the eigenvalue $n^j = \sum_{m_F = -1}^{1} n^j_{m_F}$. Then, the full Hilbert space is obtained by taking the tensor product with respect to all wells, i.e. $\mathcal{H} = \otimes_{j=1}^N \mathcal{H}^j$, such that a generic element of the full Fock basis reads
\begin{equation}
    \ket{n^1_1, n_0^1, n_{-1}^1; ...; n^N_1, n_0^N, n_{-1}^N} \equiv \bigotimes_{j = 1}^{N} \ket{n^j_1, n^j_0, n^j_{-1}}.
    \label{eq:FockStates}
\end{equation}

\subsection{Hamiltonian}
\label{subsec:FullHamiltonian}
The full Hamiltonian is composed of two main terms: a single-well Hamiltonian $\boldsymbol{H}_{\text{sw}}^j$ describing the on-site dynamics of well $j$ as well as a tunneling Hamiltonian $\boldsymbol{H}_{\text{t}}^j$ encoding the coupling between neighboring wells $j$ and $j+1$, such that in total we have
\begin{equation}
    \boldsymbol{H} = \sum_{j=1}^N \boldsymbol{H}_{\text{sw}}^j + \sum_{j=1}^{N \text{ or } N-1} \boldsymbol{H}_{\text{t}}^j.
    \label{eq:FullHamiltonian}
\end{equation}
For the sake of generality, we keep the type of boundary conditions open at this point. More precisely, we allow for periodic or open boundary conditions, which are implemented by the second sum in \eqref{eq:FullHamiltonian} running up to $N$ or $N-1$, i.e., coupling the first and the last well or not, respectively.

\begin{figure}[t!]
    \centering
    \includegraphics[width=0.9\columnwidth]{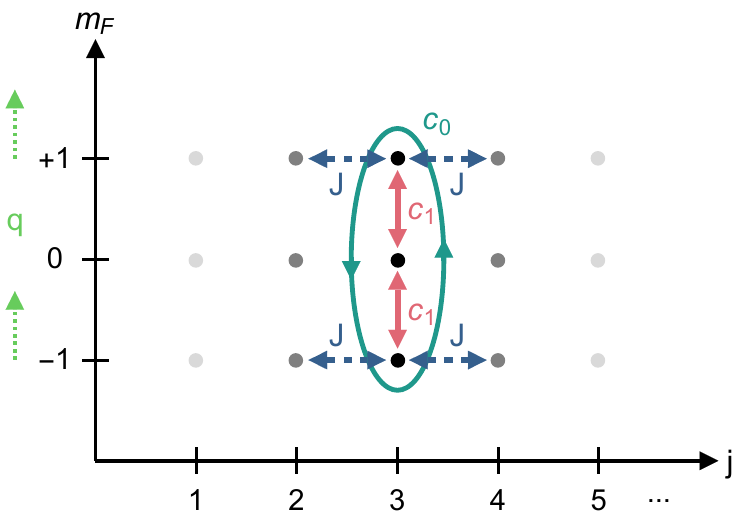}
    \caption{Illustration of the dynamics generated by the five contributions to the full Hamiltonian \eqref{eq:FullHamiltonian} for the $j=3$ well. First-order (tunneling) and second-order (collisions) processes are indicated by dashed and solid arrows, respectively, while the energy shift due to the Zeemann effect is depicted by a dotted arrow.}
    \label{fig:FullHamiltonianIllustration}
\end{figure}

The single-well Hamiltonian for well $j$ is given by
\begin{equation}
    \begin{split}
        \boldsymbol{H}^j_{\text{sw}} &= c_0 \, \boldsymbol{N}^j \left( \boldsymbol{N}^j - \mathds{1} \right) \\
        &+ c_1 \Big[ \left( \boldsymbol{N}^j_0 - (1/2) \mathds{1} \right) \left( \boldsymbol{N}^j_1 + \boldsymbol{N}^j_{-1} \right) \\
        &\hspace{1cm}+ \boldsymbol{a}_0^{j \dagger} \boldsymbol{a}_0^{j \dagger} \boldsymbol{a}_1^{j} \boldsymbol{a}_{-1}^{j} + \boldsymbol{a}_{1}^{j \dagger} \boldsymbol{a}_{-1}^{j \dagger} \boldsymbol{a}_0^{j} \boldsymbol{a}_{0}^{j} \Big] \\
        &+ q \left( \boldsymbol{N}_1^{j} + \boldsymbol{N}_{-1}^{j} \right),
    \end{split}
    \label{eq:SingleWellHamiltonian}
\end{equation}
and has been investigated in great detail, see e.g. \cite{Kawaguchi2012,Hamley2012,Kunkel2018,Kunkel2019,Kunkel2019b,Kunkel2021}.
It contains three contributions: First, the on-site density-density interaction $c_0 > 0$ describing the repulsive interaction of atoms regardless of their hyperfine levels (petrol ellipse in \autoref{fig:FullHamiltonianIllustration}). Second, the spin-changing collision interaction $c_1 < 0$ encoding the generation of spin pairs in the $m_F = \pm 1$ hyperfine levels from the $m_F=0$ level (red arrows), or the reverse process. For clarity, we omitted the mean-field shifts in the second line of Eq.~\eqref{eq:SingleWellHamiltonian}. Third, we included the parameter $q > 0$ (green arrows) representing the quadratic Zeemann shift and a possible AC Stark shift tunable via off-resonant microwave coupling between the $m_F=0$ states of the $F=1$ and $F=2$ hyperfine manifolds (mircowave dressing). In the regime of our interest, that is, low magnetic fields, the single-well couplings $c_0$ and $c_1$ are not easily tunable, and their ratio is set by the atomic species (for Lithium-7, we have $\abs{c_1 / c_0} \approx 1/2$). The parameter $q$, however, is widely tunable via the microwave dressing. 

The tunneling between equal hyperfine levels $m_F = \pm 1$ of neighboring wells $j$ and $j+1$ is described by
\begin{equation}
    \boldsymbol{H}^j_{\text{t}} = - J \sum_{m_F=\pm 1} \left( \boldsymbol{a}_{m_F}^{j \dagger} \boldsymbol{a}_{m_F}^{j + 1} + \boldsymbol{a}_{m_F}^{(j +1) \dagger} \boldsymbol{a}_{m_F}^{j} \right),
    \label{eq:TunnelHamiltonian}
\end{equation}
with a non-negative tunnel rate $J \ge 0$ (blue arrows) tunable via the strength of the optical lattice potential. Note again that the $m_F = 0$ mode does not couple to neighboring wells and also that in the limit $J \to 0$, the wells evolve independently.

\subsection{Observables: Spin operators}
\label{subsec:SpinOperators}
Among the local observables in every well are the eight spin operators that form a representation of the Lie algebra $\mathfrak{su}(3)$ \cite{Kawaguchi2012,Hamley2012}. They are constructed following the Jordan-Schwinger map \cite{Schwinger1952,Kim1987}: starting from an irreducible representation of the eight three-dimensional matrices $G_{\alpha}$ of $\mathfrak{su}(3)$ defined via
\begin{equation}
    [G_{\alpha}, G_{\beta}] = f_{\alpha \beta \gamma} G_{\gamma}
    \label{eq:LieAlgebra},
\end{equation}
where $f_{\alpha \beta \gamma}$ are the so-called structure constants (we omit their explicit form here) and $\alpha \in \{1, ..., 8\}$ labels the matrices, the local quantum operators $\boldsymbol{G}^j_{\alpha}$ obeying the very same algebra are constructed following
\begin{equation}
    \boldsymbol{G}^j_{\alpha} = \sum_{m_F, m_{F'} = -1}^{1} \boldsymbol{a}^{j \dagger}_{m_F} \left( G_{\alpha} \right)_{m_F m_{F'}} \boldsymbol{a}^j_{m_{F'}},
    \label{eq:JordanSchwingerMap}
\end{equation}
with $\left( G_{\alpha} \right)_{m_F m_{F'}}$ denoting the $(m_F, m_{F'})$-th. entry of the spin matrix $G_{\alpha}$.

A common choice for the $\mathfrak{su}(3)$ matrices $G_{\alpha}$ is made by first constructing a $\mathfrak{su}(2)$ subspace (see \cite{Hamley2012} for explicit expressions for the corresponding matrices) via
\begin{equation}
    [S^j_{\alpha}, S^j_{\beta}] = i \epsilon_{\alpha \beta \gamma} S^j_{\gamma},
\end{equation}
where $\alpha, \beta, \gamma \in \{x, y, z \}$, and thereupon deriving the three corresponding spin operators from \eqref{eq:JordanSchwingerMap}, leading to
\begin{equation}
    \begin{split}
        \boldsymbol{S}^j_x &= \frac{1}{\sqrt{2}} \left[\boldsymbol{a}_0^{j \dagger} \left( \boldsymbol{a}^j_{1} + \boldsymbol{a}^j_{-1} \right) + \left( \boldsymbol{a}^{j \dagger}_{1} + \boldsymbol{a}^{j\dagger}_{-1} \right) \boldsymbol{a}_0^{j} \right], \\
        \boldsymbol{S}^j_y &= \frac{i}{\sqrt{2}} \left[\boldsymbol{a}_0^{j \dagger} \left( \boldsymbol{a}^j_{1} - \boldsymbol{a}^j_{-1} \right) - \left( \boldsymbol{a}^{j \dagger}_{1} - \boldsymbol{a}^{j\dagger}_{-1} \right) \boldsymbol{a}_0^{j} \right], \\
        \boldsymbol{S}^j_z &= \boldsymbol{a}^{j \dagger}_{1} \boldsymbol{a}^j_{1} - \boldsymbol{a}^{j \dagger}_{-1} \boldsymbol{a}^j_{-1}.
    \end{split}
    \label{eq:SU2SpinOperators}
\end{equation}
The remaining five operators are the so-called quadrupole operators, defined as
\begin{equation}
    \boldsymbol{Q}^j_{\alpha \beta} = \{ \boldsymbol{S}^j_{\alpha}, \boldsymbol{S}^j_{\beta} \} - \frac{4}{3} \delta_{\alpha \beta} \mathds{1},
    \label{eq:QuadrupoleOperators}
\end{equation}
with $\{.,.\}$ denoting the anticommutator. Here $\boldsymbol{Q}^j_{y z}$ and $\boldsymbol{Q}^j_{xz}$ are of special interest, which read
\begin{equation}
    \begin{split}
        \boldsymbol{Q}^j_{y z} &= \frac{i}{\sqrt{2}} \left[\boldsymbol{a}_0^{j \dagger} \left( \boldsymbol{a}^j_{1} + \boldsymbol{a}^j_{-1} \right) - \left( \boldsymbol{a}^{j \dagger}_{1} + \boldsymbol{a}^{j\dagger}_{-1} \right) \boldsymbol{a}_0^{j} \right], \\
        \boldsymbol{Q}^j_{x z} &= \frac{1}{\sqrt{2}} \left[\boldsymbol{a}_0^{j \dagger} \left( \boldsymbol{a}^j_{1} - \boldsymbol{a}^j_{-1} \right) + \left( \boldsymbol{a}^{j \dagger}_{1} - \boldsymbol{a}^{j\dagger}_{-1} \right) \boldsymbol{a}_0^{j} \right].
    \end{split}
    \label{eq:QuadrupoleOperators2}
\end{equation}

\subsection{Readout schemes}
\label{subsec:ReadoutScheme}
We will analyze the system by gathering information about the measurement distributions of the variable pairs $(\boldsymbol{S}_x^j, \boldsymbol{Q}^j_{yz})$ over multiple wells. We discuss methods for reading out two types of such distributions in the following.

\subsubsection{Separate detection}
\label{sec:separate_detection}

In BEC experiments, the population of each mode, $n^j_{m_F}=\langle\boldsymbol{N}^j_{m_F}\rangle$ is detected through absorption imaging after Stern-Gerlach separation of the different $m_F$ components. From this the $z$-component of the local spin can be extracted as $S^j_z=n^j_{1}-n^j_{-1}$.
Other spin components can be measured by applying a radio-frequency (rf) magnetic field prior to the absorption imaging. For a frequency matching the linear Zeeman shift, resonant Rabi oscillations are driven, described by the Hamiltonian (in the rotating wave approximation) \cite{Kunkel2018,Kunkel2019b}
\begin{equation}
    \boldsymbol{H}_{\rm rf}=\sum_{j=1}^N \Omega_{\rm rf}\left[\cos(\phi_{\rm rf})\boldsymbol{S}_y^j - \sin(\phi_{\rm rf})\boldsymbol{S}_x^j \right]
\end{equation}
Here, $\Omega_{\rm rf}$ is the Rabi frequency of the drive, and $\phi_{\rm rf}$ is a tunable phase. Applying the drive for a time $t=\pi/(2\Omega_{\rm rf})$ thus allows one to map the spin along an arbitrary direction of the equatorial plane of the spin sphere onto $\boldsymbol{S}_z^j$.
In the case where the $m_F=0$ mode is a coherent state with a population far larger than the side modes, this scheme is analogous to homodyne detection in optics, where the signal mode(s) are mixed with a local oscillator mode on a beam splitter to extract the field quadratures \cite{Gross2011}.

To toggle between measurements of spin operators and quadrupole operators, one can add a time delay before the application of the rf-rotation. By tuning the quadratic Zeeman shift and the microwave dressing field in the single-well Hamiltonian Eq.~\eqref{eq:SingleWellHamiltonian}, the $m_F=0$ mode acquires a relative phase. For instance, measuring $\boldsymbol{Q}_{yz}^j$ instead of $\boldsymbol{S}_x^j$ requires an additional phase of $\pi/2$.

\subsubsection{Simultaneous detection}
\label{sec:simultaneous_detection}
Sophisticated methods to extract joint distributions over $(\boldsymbol{S}_x^j, \boldsymbol{Q}^j_{yz})$, or equivalently $(\boldsymbol{S}_y^j, \boldsymbol{Q}^j_{xz})$, in close analogy to the heterodyne detection protocol in quantum optics \cite{Schleich2001,Weedbrook2012,Mandel2013,Serafini2017} have been experimentally realized rather recently \cite{Kunkel2019,Kunkel2019b,Kunkel2021}. For this, one exploits the availability of additional, initially unoccupied, internal levels, in our example, the $F=2$ hyperfine manifold with five additional Zeeman levels. Using microwave fields resonantly coupling the levels of the $F=1$ manifold with those of the $F=2$ manifold, one can realize analogs of beam splitter operations between them. Splitting each mode equally and subsequently applying different rf-rotations in the two manifolds realizes the simultaneous detection of, for example, $\boldsymbol{S}_x^j$ and $\boldsymbol{Q}^j_{yz}$ (see Ref.~\cite{Kunkel2019} for details). In the case of the $m_F=0$ mode being macroscopically occupied, this corresponds to sampling from the Husimi $Q$-distribution in the phase space spanned by $\boldsymbol{S}_x^j$ and $\boldsymbol{Q}^j_{yz}$ (see \autoref{sec:ContinousVariables} for details).

\section{From a spin-1 BEC to a continuous-variable quantum system}
\label{sec:ContinousVariables}
We point out the connection between the system of our interest, a multi-well spin-1 BEC, and a continuous-variable quantum system describing coupled bosonic oscillator modes. Thereupon, we introduce several kinds of phase-space descriptions.

\subsection{Undepleted pump approximation}
\label{subsec:UndepletedPumpApproximation}
The undepleted pump regime captures the early-time dynamics under the Hamiltonian \eqref{eq:FullHamiltonian} when initially preparing the polar ground state of the spin-1 BEC in every well, i.e.
\begin{equation}
\label{eq:PolarState}
\ket{\psi(0)}=\ket{0, \alpha, 0}^{\otimes N}
\end{equation}
where $\ket{\alpha}$ is the local coherent state of the zero mode with mean particle number $n=|\alpha|^2$. For large values of $n$ the $m_F = 0$ mode is macroscopically populated compared to the side modes $m_F = \pm 1$ in every well, i.e.
\begin{equation}
    \braket{\boldsymbol{N}^j_0} \gg \braket{\boldsymbol{N}^j_{\pm 1}}
    \label{eq:UndepletedPumpApproximation},
\end{equation}
for all $j$, which is equivalent to 
\begin{equation}
    n^j = \braket{\boldsymbol{N}^j} \approxeq \braket{\boldsymbol{N}_0^j}.
\end{equation}
Under the above assumptions, we can approximate the $m_F = 0$ mode operators by their norms
\begin{equation}
    \boldsymbol{a}_0^j = \boldsymbol{a}_0^{j \dagger} \approxeq \sqrt{n^j},
    \label{eq:UndepletedPumpApproximationZeroMode}
\end{equation}
such that $\boldsymbol{N}_0^j \approxeq n^j$. Then, the full Hamiltonian \eqref{eq:FullHamiltonian} simplifies to
\begin{align}
    \begin{split}
        \boldsymbol{H}_{\text{up}} &= \sum_{j=1}^{N} \Bigg[ c_0 \Big( \boldsymbol{a}_{1}^{j \dagger} \boldsymbol{a}_{1}^{j \dagger} \boldsymbol{a}_{1}^{j} \boldsymbol{a}_{1}^{j} + \boldsymbol{a}_{-1}^{j \dagger} \boldsymbol{a}_{-1}^{j \dagger} \boldsymbol{a}_{-1}^{j} \boldsymbol{a}_{-1}^{j} + 2 \boldsymbol{N}_{1}^{j} \boldsymbol{N}_{-1}^{j} \Big) \\
        &\hspace{1.2cm}+ \tilde{c}^j_1 \left( \boldsymbol{a}_1^{j} \boldsymbol{a}_{-1}^{j} + \boldsymbol{a}_{1}^{j \dagger} \boldsymbol{a}_{-1}^{j \dagger} \right) \\
        &\hspace{1.2cm}+ \left(\tilde{q}^j + c_0 n^j \right) \left( \boldsymbol{N}_1^{j} + \boldsymbol{N}_{-1}^{j} \right) \Bigg] \\
        &-J \sum_{j=1}^{N \text{ or } N-1} \Big( \boldsymbol{a}_{-1}^{j \dagger} \boldsymbol{a}_{-1}^{j + 1} + \boldsymbol{a}_{-1}^{(j +1) \dagger} \boldsymbol{a}_{-1}^{j} \\
        &\hspace{2.3cm}+ \boldsymbol{a}_{1}^{j \dagger} \boldsymbol{a}_{1}^{j + 1} + \boldsymbol{a}_{1}^{(j +1) \dagger} \boldsymbol{a}_{1}^{j} \Big) \\
        &+ H_{\text{offset}}.
    \end{split}
    \label{eq:TotalHamiltonianUndepletedPump}
\end{align}
Here, we defined the rescaled couplings
\begin{equation}
    \tilde{c}^j_1 = c_1 n^j, \quad \tilde{q}^j = c_1 \left(n^j - \frac{1}{2} \right) + q, 
    \label{eq:RescaledCouplings}
\end{equation}
as well as a constant offset term
\begin{equation}
    H_{\text{offset}} = c_0 \sum_{j=1}^N (n^j)^2,
\end{equation}
which we will drop in the following.

\subsection{Relative modes and canonically conjugate variables}
\label{subsec:RelativeModes}
In the undepleted pump regime, the relevant degrees of freedom and their Hamiltonian \eqref{eq:TotalHamiltonianUndepletedPump} can be mapped to a continuous-variable quantum system. To that end, we introduce the relative mode operators between the side modes
\begin{equation}
    \boldsymbol{a}^j_{\pm} = \frac{1}{\sqrt{2}} \left( \boldsymbol{a}_{1}^j \pm \boldsymbol{a}^j_{-1} \right),
    \label{eq:RelativeModesOperators}
\end{equation}
which also represent independent bosonic modes, since \eqref{eq:BosonicCommutationRelations} implies
\begin{equation}
    [\boldsymbol{a}_{\pm}^j, \boldsymbol{a}_{\pm}^{j' \dagger}] = \delta_{j j'}, \quad [\boldsymbol{a}_{\pm}^j, \boldsymbol{a}_{\mp}^{j' \dagger}] = 0.
    \label{eq:RelativeModesCommutator}
\end{equation}
Their associated canonical operators are defined as
\begin{equation}
    \boldsymbol{\phi}_{\pm}^j = \frac{1}{\sqrt{2}} \left( \boldsymbol{a}^{j \dagger}_{\pm} + \boldsymbol{a}^j_{\pm} \right), \quad \boldsymbol{\pi}_{\pm}^j = \frac{i}{\sqrt{2}} \left( \boldsymbol{a}^{j \dagger}_{\pm} - \boldsymbol{a}^j_{\pm} \right),
    \label{eq:CanonicalOperators}
\end{equation}
which fulfill the canonical commutation relations
\begin{equation}
    \left[ \boldsymbol{\phi}^j_{\pm}, \boldsymbol{\pi}^{j'}_{\pm} \right] = i \delta^{j j'}, \quad \left[ \boldsymbol{\phi}^j_{\pm}, \boldsymbol{\pi}^{j'}_{\mp} \right] = 0.
    \label{eq:CanonicalOperatorsCommutator}
\end{equation}
Interestingly, in the undepleted pump regime, the two pairs of canonical operators are equivalent to pairs of spin operators up to normalization, to wit
\begin{equation}
    \begin{split}
        \boldsymbol{S}^j_x &= \sqrt{2 n^j} \boldsymbol{\phi}^j_+, \quad \boldsymbol{Q}^j_{yz} = - \sqrt{2 n^j} \boldsymbol{\pi}^j_+, \\
        \boldsymbol{S}^j_y &= \sqrt{2 n^j} \boldsymbol{\phi}^j_-, \quad \boldsymbol{Q}^j_{xz} = - \sqrt{2 n^j} \boldsymbol{\pi}^j_-,
    \end{split}
    \label{eq:RelationSpinOperatorsConjugateOperators}
\end{equation}
which follows from using \eqref{eq:UndepletedPumpApproximationZeroMode} in \eqref{eq:SU2SpinOperators} and \eqref{eq:QuadrupoleOperators2}.
This shows that, for short times, the relevant degrees of freedom mimic two pairs of canonical variables, and hence the local Hilbert spaces $\mathcal{H}^j$ decompose as $\mathcal{H}^j = \mathcal{H}_+^j \otimes \mathcal{H}^j_-$ 

\begin{figure}[t!]
    \centering
    \includegraphics[width=0.9\columnwidth]{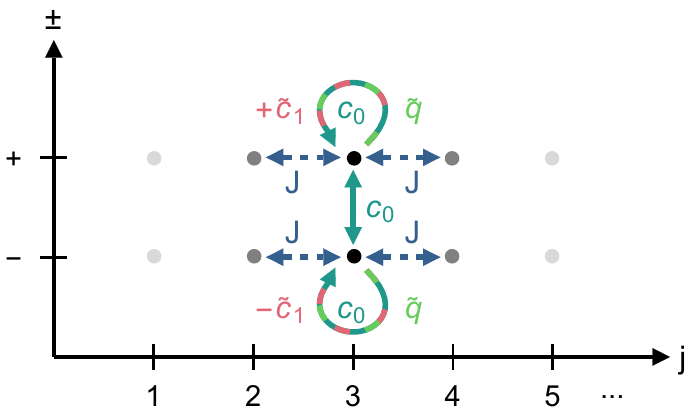}
    \caption{Sketch of the dynamics in the undepleted pump regime generated by \eqref{eq:HamiltonianUndepletedPumpDecomposition} (in analogy to \autoref{fig:FullHamiltonianIllustration}).}
    \label{fig:UPHamiltonianIllustration}
\end{figure}

Next, we express the Hamiltonian \eqref{eq:TotalHamiltonianUndepletedPump} in terms of the newly defined operators (see \hyperref[app:OperatorIdentitiesUndepletedPump]{Appendix} \ref{app:OperatorIdentitiesUndepletedPump} for details). We find the general decomposition
\begin{equation}
    \boldsymbol{H}_{\text{up}} = \boldsymbol{H}_{\text{up}}^{+} + \boldsymbol{H}_{\text{up}}^{-} + \boldsymbol{H}_{\text{up}}^{\text{mix}},
    \label{eq:HamiltonianUndepletedPumpDecomposition}
\end{equation}
where the form-equivalent $\boldsymbol{H}_{\text{up}}^{+}$ and $\boldsymbol{H}_{\text{up}}^{-}$ contain all terms with only $+$ or $-$ modes, respectively, while $\boldsymbol{H}_{\text{up}}^{\text{mix}}$ contains all terms mixing the two relative modes $\pm$. In terms of relative mode operators, these two types of Hamiltonians read
\begin{equation}
    \begin{split}
        \boldsymbol{H}_{\text{up}}^{\pm} &= \sum_{j=1}^N \Bigg[ c_0 \, \boldsymbol{a}_{\pm}^{j \dagger} \boldsymbol{a}_{\pm}^{j} \boldsymbol{a}_{\pm}^{j \dagger} \boldsymbol{a}_{\pm}^{j} \\
        &\hspace{1.2cm}+ \left[ \tilde{q}^j + c_0 \left( n^j - 1 \right) \right] \boldsymbol{N}^j_{\pm} \\
        &\hspace{1.2cm}\pm \frac{\tilde{c}^j_1}{2} \left( \boldsymbol{a}_{\pm}^{j \dagger} \boldsymbol{a}_{\pm}^{j \dagger} + \boldsymbol{a}_{\pm}^{j} \boldsymbol{a}_{\pm}^{j} \right) \Bigg] \\
        &- J \sum_{j=1}^{N \text{ or } N-1} \left( \boldsymbol{a}_{\pm}^{j \dagger} \boldsymbol{a}_{\pm}^{j + 1} + \boldsymbol{a}_{\pm}^{(j +1) \dagger} \boldsymbol{a}_{\pm}^{j} \right), \\
        \boldsymbol{H}_{\text{up}}^{\text{mix}} &= 2 c_0 \sum_{j=1}^N \boldsymbol{N}_+^{j} \boldsymbol{N}_-^{j},
    \end{split}
    \label{eq:HamiltonianUndepletedPumpRelativeModes}
\end{equation}
whose dynamics are sketched in \autoref{fig:UPHamiltonianIllustration}. From the latter formulas, it becomes apparent that in the undepleted pump regime, the relative mode operators, or equivalently, the two pairs of canonical operators, constitute a complete set of observables for characterizing the early-time dynamics. Further, the dynamics within the $\pm$ phase spaces differ only by a sign in the $\pm \tilde{c}_1$ term.

\subsection{Canonical phase-space}
\label{subsec:CanonicalPhaseSpace}
As the observables of our interest now correspond to continuous variables, we can faithfully apply the powerful concepts of phase-space descriptions. The two sets of canonical operators $(\boldsymbol{\phi}^j_{\pm}, \boldsymbol{\pi}^j_{\pm})$ span the two two-dimensional canonical phase spaces for well $j$, which are known to be isomorphic to the Euclidean plane $\mathbb{R}^2$ with measure \cite{Zhang1990,Bengtsson2017}
\begin{equation}
    \int \mathrm{d}\phi_{\pm}^j \, \mathrm{d}\pi_{\pm}^j.
    \label{eq:PhaseSpaceMeasure}
\end{equation}
The field operators are conveniently combined into a single vector $\boldsymbol{\chi}_{\pm}^j$ by defining $\boldsymbol{\chi}_{\pm}^j = \boldsymbol{\phi}_{\pm}^j$ for $j \in [1,N]$ and $\boldsymbol{\chi}_{\pm}^j = \boldsymbol{\pi}^{j-N}_{\pm}$ for $j \in [N+1,2N]$, which we formally write as \cite{Weedbrook2012,Serafini2017}
\begin{equation}
    \boldsymbol{\chi}_{\pm} = ( \boldsymbol{\phi}_{\pm}, \boldsymbol{\pi}_{\pm})^T = (
    \boldsymbol{\phi}_{\pm}^1, \dots, \boldsymbol{\phi}_{\pm}^N, \boldsymbol{\pi}_{\pm}^1, \dots, \boldsymbol{\pi}_{\pm}^N)^T.
\end{equation}
The canonical commutation relations \eqref{eq:CanonicalOperatorsCommutator} become
\begin{equation}
    [\boldsymbol{\chi}_{\pm}^j, \boldsymbol{\chi}_{\pm}^{j'}] = i \Omega^{j j'} \boldsymbol{\mathds{1}},
    \label{eq:GroupedOperatorsCommutator}
\end{equation}
where
\begin{equation}
    \Omega = (i \sigma_2)  \otimes \mathds{1}_N
\end{equation}
denotes the symplectic metric revealing the symplectic structure of the canonical phase space, and $\sigma_2$ is the second Pauli matrix.

For every well $j$, we define the set of canonical coherent states associated with the relative modes $\pm$ as displaced vacuum states \cite{Schleich2001,Weedbrook2012,Mandel2013,Serafini2017}, i.e.
\begin{equation}
    \ket{\alpha^j_{\pm}} = \boldsymbol{D} (\alpha^j_{\pm}) \ket{0^j_{\pm}},
    \label{eq:CanonicalCoherentStates}
\end{equation}
where $\boldsymbol{D} (\alpha^j_{\pm})$ is the unitary displacement operator
\begin{equation}
    \boldsymbol{D} (\alpha^j_{\pm}) = e^{\alpha^j_{\pm} \boldsymbol{a}^{j \dagger}_{\pm} - \alpha^{*j}_{\pm} \boldsymbol{a}^{j}_{\pm}},
    \label{eq:DisplacementOperator}
\end{equation}
with the complex-valued phase fields being parameterized in terms of cartesian coordinates as
\begin{equation}
    \alpha^j_{\pm} = \frac{1}{\sqrt{2}} \left( \phi_{\pm}^j + i \pi_{\pm}^j \right).
    \label{eq:PhaseFields}
\end{equation}
Importantly, the set of coherent states constitutes an overcomplete basis, i.e., the coherent states defined in \eqref{eq:CanonicalCoherentStates} resolve the identity in the $\pm$ subspaces
\begin{equation}
    \mathds{1} = \int \frac{\mathrm{d} \phi_{\pm}^j \pi_{\pm}^{j}}{2 \pi} \ket{\boldsymbol{\alpha}_{\pm}^j} \bra{\boldsymbol{\alpha}_{\pm}^j},
    \label{eq:CoherentStateIdentity}
\end{equation}
but are not orthogonal to each other.

\subsection{Phase space distributions}
\label{subsec:PhaseSpaceDistributions}
The phase-space picture enables the description of the system's state $\boldsymbol{\rho}^j$ in terms of classical phase-space distributions. In the following, we will consider the distributions associated with either the $+$ or the $-$ mode, whose corresponding density operators are obtained via the partial trace $\boldsymbol{\rho}^j_{\pm} = \Tr_{\mp} \{ \boldsymbol{\rho}^j \}$. 

\subsubsection{Wigner $W$-distribution}
The arguably most prominent phase-space representation is the Wigner $W$-distribution \cite{Wigner1932}, which is defined as the Fourier transform of the characteristic function, namely \cite{Weedbrook2012}
\begin{equation}
    \begin{split}
        \mathcal{W}^j_{\pm} &\equiv \mathcal{W}^j_{\pm} (\phi_{\pm}^j, \pi_{\pm}^j) \\
        &= \int \frac{\mathrm{d} \tilde{\phi}_{\pm}^{j} \, \mathrm{d} \tilde{\pi}_{\pm}^{j}}{2 \pi} \, e^{-i (\phi_{\pm}^j, \pi_{\pm}^j) \Omega (\tilde{\phi}_{\pm}^j, \tilde{\pi}_{\pm}^j)^T} \\
        &\hspace{0.9cm} \times \Tr \left\{ \boldsymbol{\rho}^j_{\pm} \, e^{i (\boldsymbol{\phi}_{\pm}^j, \boldsymbol{\pi}_{\pm}^j) \Omega (\tilde{\phi}_{\pm}^j, \tilde{\pi}_{\pm}^j)^T} \right\}.
    \end{split}
    \label{eq:WignerWDistribution}
\end{equation}
Here, we have chosen the normalization such that $\mathcal{W}^j_{\pm}$ is normalized to unity with respect to the phase-space integral measure \eqref{eq:PhaseSpaceMeasure}, i.e.
\begin{equation}
    1 = \Tr \left\{ \boldsymbol{\rho}^j_{\pm} \right\} = \int \mathrm{d} \phi_{\pm}^j \, \mathrm{d}\pi_{\pm}^j \, \mathcal{W}^j_{\pm}.
\end{equation}
It is well-known that the Wigner $W$-distribution can become negative when the underlying state is non-classical \cite{Kenfack2004}. However, we note that both our simulation approaches are naturally based on positive Wigner $W$-distributions (see \autoref{sec:Dynamics}).

\subsubsection{Marginal distributions}
Next, we introduce the measurement distributions over the canonical operators
\begin{equation}
    \begin{split}
        f^j_{\pm} &\equiv f^j_{\pm} (\phi^j_{\pm}) = \Tr \left\{ \boldsymbol{\rho}^j_{\pm} \ket{\boldsymbol{\phi}^j_{\pm}} \bra{\boldsymbol{\phi}^j_{\pm}} \right\}, \\
        g^j_{\pm} &\equiv g^j_{\pm} (\pi^j_{\pm}) = \Tr \left\{ \boldsymbol{\rho}^j_{\pm} \ket{\boldsymbol{\pi}^j_{\pm}} \bra{\boldsymbol{\pi}^j_{\pm}} \right\},
    \end{split}
    \label{eq:MarginalDistributions}
\end{equation}
with the eigenvalue equations
\begin{equation}
    \boldsymbol{\phi}^j_{\pm} \ket{\phi^j_{\pm}} = \phi^j_{\pm} \ket{\phi^j_{\pm}}, \quad \boldsymbol{\pi}^j_{\pm} \ket{\pi^j_{\pm}} = \pi^j_{\pm} \ket{\pi^j_{\pm}},
\end{equation}
understood. The distributions \eqref{eq:MarginalDistributions} are true probability density functions with normalizations
\begin{equation}
    1 = \Tr \left\{ \boldsymbol{\rho}^j_{\pm} \right\} = \int \mathrm{d} \phi_{\pm}^j \, f^j_{\pm} = \int \mathrm{d} \pi_{\pm}^j \, g^j_{\pm},
\end{equation}
and can be identified with the marginals of the Wigner $W$-distributions
\begin{equation}
    f^j_{\pm} = \int \mathrm{d}\pi_{\pm}^j \, \mathcal{W}^j_{\pm}, \quad g^j_{\pm} = \int \mathrm{d}\phi_{\pm}^j \, \mathcal{W}^j_{\pm}.
\end{equation}
As such, they encode the full information of $\phi^j_{\pm}$ and $\pi^j_{\pm}$, but no information about their correlations. Hence, in the special case when $\phi^j_{\pm}$ and $\pi^j_{\pm}$ are uncorrelated, the corresponding Wigner $W$-distribution decomposes into a product $W^j_{\pm} = f^{j}_{\pm} \, g^j_{\pm}$.

The separate detection scheme, or homodyne detection \cite{Braunstein1990}, described in \autoref{sec:separate_detection}, approximately corresponds to sampling from the marginal distributions. Fundamentally, of course, this cannot generally be an exact correspondence for any finite atom number, as the detected quantities $\boldsymbol{n}^j_{m_F}$ have a discrete spectrum (detected atom numbers), while the desired Wigner marginals are continuous. However, in the undepleted pump approximation, the Wigner marginals are generally well-approximated by the homodyne statistics \cite{Kunkel2019b, Kunkel2018}.

\subsubsection{Husimi $Q$-distribution}
Since the coherent state projectors $\ket{\boldsymbol{\alpha}_{\pm}^j} \bra{\boldsymbol{\alpha}_{\pm}^j}$ are non-negative and resolve the identity, they constitute a positive operator-valued measure (POVM). The corresponding measurement distribution is the so-called Husimi $Q$-distribution \cite{Husimi1940}
\begin{equation}
    \mathcal{Q}^j_{\pm} \equiv \mathcal{Q}^j_{\pm} (\phi_{\pm}^j, \pi_{\pm}^j) = \Tr \left\{ \boldsymbol{\rho}^j_{\pm} \ket{\boldsymbol{\alpha}^j_{\pm}} \bra{\boldsymbol{\alpha}^j_{\pm}} \right\}.
    \label{eq:HusimiQDistribution}
\end{equation}
While the Wigner $W$-distribution can be negative, the Husimi $Q$-distribution is \textit{always} non-negative \cite{Cartwright1976} since it stems from a POVM. More specifically, it is bounded by $0 \le \mathcal{Q}^j_{\pm} \le 1$ \cite{Schleich2001,Mandel2013}. The normalization is induced by \eqref{eq:CoherentStateIdentity}, leading to
\begin{equation}
    1 = \Tr \left\{ \boldsymbol{\rho}^j_{\pm} \right\} = \int \frac{\mathrm{d} \phi_{\pm}^j \, \mathrm{d}\pi_{\pm}^j}{2 \pi} \, \mathcal{Q}^j_{\pm}.
\end{equation}

When operating in the regime $n^j_0 \gg n^j_{\pm1}$ the simultaneous detection scheme described in \autoref{sec:simultaneous_detection} can be identified with the so-called eight-port-homodyne detection in quantum optics (also referred to as heterodyne detection). In this scheme, each measurement corresponds to drawing a sample from the Husimi-Q-distribution of the signal field \cite{Leonhardt1995}.

\section{Simulating the Dynamics}
\label{sec:Dynamics}
We simulate the dynamics of the spin-1 BEC in two different approximations. To capture most of the full Hamiltonian and to check the validity of the undepleted pump approximation, we perform a truncated Wigner simulation, see \autoref{subsec:TWA}. Additionally, we derive an analytically solvable model based on \eqref{eq:HamiltonianUndepletedPumpDecomposition} in \autoref{subsec:GaussianModel}.

\subsection{Truncated Wigner Approximation}
\label{subsec:TWA}
In the regime of high occupations, it is usually infeasible to exactly solve the exponentially complex dynamics in the full Hilbert space $\mathcal{H}$. Instead, as the populations increase, quantum fluctuations compared to the mean-field dynamics become less relevant, and hence, the predictions made by semiclassical approximation methods become more and more accurate. One such semiclassical technique is the truncated Wigner approximation (TWA). At its core, mode operators (for brevity, we drop indices in this subsection) are replaced by complex numbers \cite{Polkovnikov2010}
\begin{equation}
    \boldsymbol{a}^{(\dagger)} \to \alpha^{(*)},
    \label{eq:TWAOperatorsToCNumbers}
\end{equation}
which amounts to a lowest-order expansion of the Wigner-Weyl correspondence rules. Consequently, the Hamiltonian operator $\boldsymbol{H}(\boldsymbol{a}, \boldsymbol{a}^{\dagger})$ reduces to a classical Hamiltonian function $H(\alpha, \alpha^*)$ of the complex-valued phase-space coordinates $(\alpha,\alpha^{*})$.

In general, the von Neumann equation
\begin{equation}
    i\partial_t \boldsymbol{O}(t) = [\boldsymbol{O}(t),\boldsymbol{H}],
    \label{eq:HeisenbergPicture}
\end{equation}
for some time-dependent operator $\boldsymbol{O}$, can be translated into an equation of motion for the operator's Weyl symbol $O$ in phase space via the Moyal bracket
\begin{equation}
    \{A, B\}_{\text{MB}} = \frac{A}{2} \sin \left[ 2 \left(\overleftarrow{\partial}_{\alpha} \overrightarrow{\partial}_{\alpha^*} - \overleftarrow{\partial}_{\alpha^*} \overrightarrow{\partial}_{\alpha} \right)\right] B,
    \label{eq:MoyalBracketDef}
\end{equation}
which results in \cite{Polkovnikov2010}
\begin{equation}
    i\partial_t O(t) = \{O(t), H\}_{\text{MB}}.
    \label{eq:EOMMoyalBracket}
\end{equation}
In TWA, one is interested in the evolution of the elementary $c$-numbers $\alpha$ representing the mode operators to leading order, for which Eq.~\eqref{eq:EOMMoyalBracket} simplifies to the classical Poisson bracket
\begin{equation}
    i\partial_t \alpha (t) = \partial_{\alpha^* (t)} H,
    \label{eq:EOMMoyalBracketSimplified}
\end{equation}
since all higher-order derivatives vanish. 

In the case when multiple modes are coupled as for the Hamiltonian of interest in Eq.~\eqref{eq:FullHamiltonian}, Eq.~\eqref{eq:EOMMoyalBracketSimplified} defines a system of coupled differential equations that may be solved numerically using a suited integrator. Before being propagated in time, the initial set of coordinates $\alpha (0)$ of a given well $j$ and mode $m_F$ is obtained as Monte Carlo samples from the initial Wigner $W$-distribution, for which two cases are relevant. When considering the vacuum $\ket{0}$, the corresponding Wigner $W$-distribution is of Gaussian form
\begin{equation}
    \label{eq:WignerSamplesVacuum}
    \ket{\psi}=\ket{0} \iff \alpha \sim \mathcal{N} \left[0, \frac{1}{2} \right] + i \mathcal{N} \left[0, \frac{1}{2} \right].
\end{equation}
The samples of coherent states $\ket{\gamma}$
have the same variance but are displaced by the square root of their mean particle number $\langle \boldsymbol{N} \rangle = |\gamma|^2$ from the origin
\begin{equation}
    \label{eq:WignerSamplesCoherent}
    \ket{\psi}=\ket{\gamma} \iff \alpha \sim \mathcal{N} \left[\mathrm{Re}(\gamma), \frac{1}{2} \right] + i \mathcal{N} \left[\mathrm{Im}(\gamma), \frac{1}{2} \right],
\end{equation}
since $\ket{\gamma}$ itself is obtained by acting with the displacement operator $\boldsymbol{D}(\gamma)$ on the vacuum, see \eqref{eq:CanonicalCoherentStates}.

In the presence of thermal fluctuations, the pure vacuum state $\ket{0}$ is replaced by the thermal ensemble $\boldsymbol{\rho} \propto \exp\left(-\beta \boldsymbol{a}^\dagger \boldsymbol{a}\right)$ with the inverse temperature $\beta = 1/T$. Accordingly, the standard deviations of the Wigner $W$-distribution are rescaled as
\begin{equation}
    \label{eq:WignerSamplesVacuumThermal}
    \frac{1}{2} \rightarrow \sqrt{1 + 2 n_{\text{BE}}(\beta)} \, \frac{1}{2},
\end{equation}
which is equivalent to adding $n_{\text{BE}}(\beta)$ to both variances, where 
\begin{equation}
    \label{eq:BoseEinsteinOccupation}
    n_{\text{BE}}(\beta) = \frac{1}{e^\beta - 1}
\end{equation}
denotes the Bose-Einstein distribution.

Then, the expectation value of some observable $\boldsymbol{O}$ is obtained as the stochastic average over all generated samples
\begin{equation}
    \langle \boldsymbol{O} \rangle_{\mathrm{TWA}} = \frac{1}{|\mathcal{S}|}\sum_{\alpha \in \mathcal{S}} O\left(\alpha, \alpha^*\right),
    \label{eq:TWAExpectationValue}
\end{equation}
where $\mathcal{S}$ denotes the set of all samples. Note here that we again rely on the correspondence between $\boldsymbol{O}$ and $O$ given in Eq.~\eqref{eq:TWAOperatorsToCNumbers}.

Having introduced all necessary tools in TWA, we want to briefly comment on its regimes of applicability. In the limit of high occupations $n \gg 1$ and initial states with minimal fluctuations, i.e., coherent states, the operator to $c$-number correspondence Eq.~\eqref{eq:TWAOperatorsToCNumbers} is a justified simplification since the relative fluctuations scale as $1/\sqrt{n}$. The fluctuations of the initial state are herein represented accurately, as they correspond to exact samples of the Wigner $W$-distribution of the initial state $\ket{\psi}$. In contrast, only those fluctuations that build up during the evolution due to the unitary evolution under $\boldsymbol{H}$ are not captured precisely. In the limit $\hbar \to 0$, the dynamics generated by TWA become exact. For a more thorough picture of this matter, see Ref. \cite{Oliva2018}, in which the impossibility of quantum phase-space trajectories for generic (i.e., anharmonic) quantum systems is discussed. For further reading regarding TWA and other semiclassical techniques, we refer the reader to \cite{Polkovnikov2010} and \cite{Blakie2008}.

\subsection{Gaussian Model}
\label{subsec:GaussianModel}
Interestingly, the short-time dynamics can be described by a simple Gaussian model to a reasonable extent, which we discuss next.

\subsubsection{Gaussian Hamiltonian}
\label{subsubsec:GaussianHamiltonian}
Starting from the undepleted pump Hamiltonian \eqref{eq:HamiltonianUndepletedPumpDecomposition}, we make two simplifying assumptions to arrive at a quadratic model: First, we neglect the atomic collisions between the atoms, i.e. set $c_0 \approxeq 0$, which amounts to dropping all fourth- and one second-order contribution in \eqref{eq:HamiltonianUndepletedPumpRelativeModes}. This results in a vanishing mixing Hamiltonian $\boldsymbol{H}_{\text{up}}^{\text{mix}} \approxeq 0$, which disentangles the $\pm$ phase spaces and hence allows us to study their dynamics independently. Second, despite no explicit translational invariance, an approximate invariance exists in the center of the lattice. Since boundary effects are negligible for the subsystem of our interest (see \autoref{subsec:QuantumEntropies}), we can assume all parameters to be equal across all wells, i.e. $\tilde{q}^j \approxeq \tilde{q}, \tilde{c}^j_1 \approxeq \tilde{c}_1$ and $n^j \approxeq n$. 

\begin{figure}[t!]
    \centering
    \includegraphics[width=0.9\columnwidth]{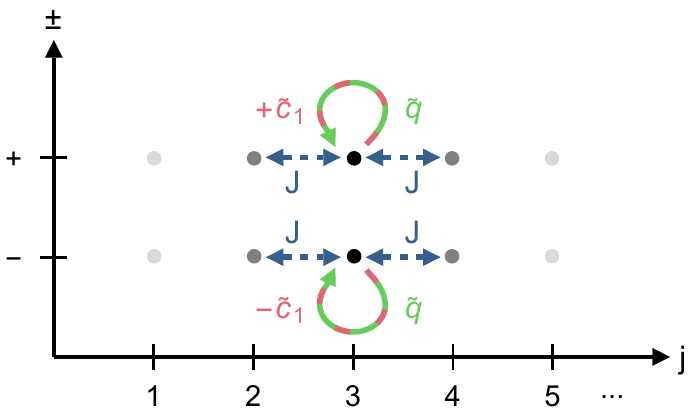}
    \caption{Sketch of the dynamics in the Gaussian regime generated by \eqref{eq:HamiltonianGaussianRelativeModes} (in analogy to \autoref{fig:FullHamiltonianIllustration}).}
    \label{fig:GaussianHamiltonianIllustration}
\end{figure}

\begin{figure*}[t!]
    \centering
    \includegraphics[width=0.95\textwidth]{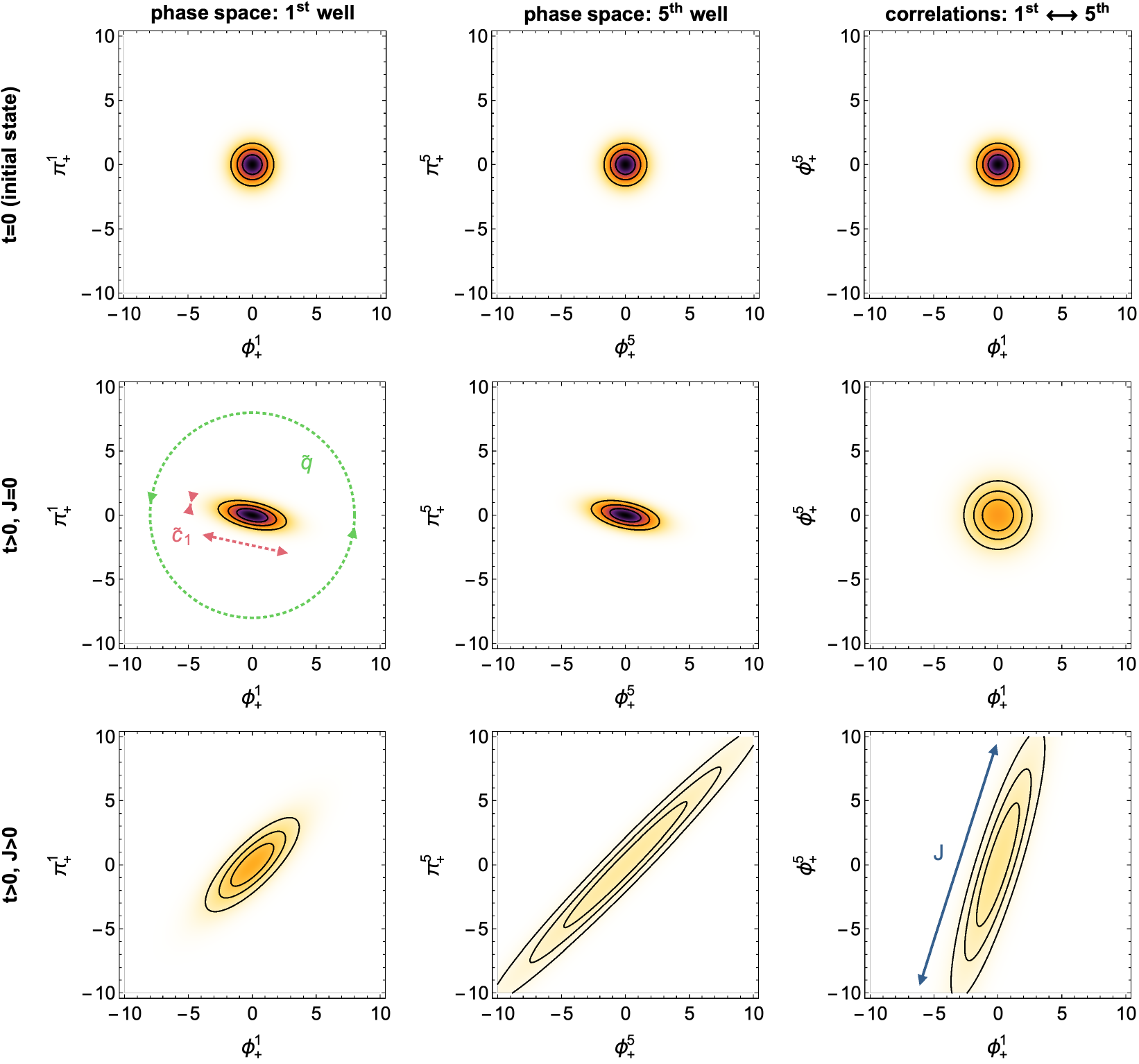}
    \caption{Illustrations of the effects of the three terms in the Gaussian Hamiltonian \eqref{eq:HamiltonianGaussianRelativeModes} on the $+$ phase-space distributions for $N=9$ wells and open boundary conditions. The columns depict the phase-space distributions of the first well (left column), the fifth well in the center (middle column), and the correlations in the fields between these two wells (right column), respectively. At $t=0$ (upper row), all wells are uncorrelated and in the same state. Hence, the distribution's covariance matrices are proportional to the identity. When the uncoupled system ($J=0$) has evolved (middle row), all local distributions are equally squeezed (red dashed arrows) and rotated (green dashed arrow) by $\tilde{c}_1$ and $\tilde{q}$, respectively, and remain uncorrelated. For a non-zero tunnel rate $J > 0$ (lower row), the local distributions are stretched proportionally (note that the stretching is equal across the wells for periodic boundary conditions), such that correlations between the wells build up, see the lower right plot (blue solid arrow).}
    \label{fig:GaussianDistributionsIllustration}
\end{figure*}

The resulting Hamiltonian is Gaussian, i.e., of second order in the relative mode operators, and reads for the $+$ mode (the $-$ mode can be treated on equal footing)
\begin{equation}
    \begin{split}
        \boldsymbol{H}_{\text{up,Gauss}}^+ &= \sum_{j=1}^N \left[ \tilde{q} \boldsymbol{N}_+^j + \frac{\tilde{c}_1}{2} \left( \boldsymbol{a}_+^j \boldsymbol{a}_+^j + \boldsymbol{a}_+^{j \dagger} \boldsymbol{a}_+^{j \dagger} \right) \right] \\
        &+ J \sum_{j=1}^{N \text{ or } N-1} \left( \boldsymbol{a}_+^{j \dagger} \boldsymbol{a}_+^{j + 1} + \boldsymbol{a}_+^{(j+1) \dagger} \boldsymbol{a}_+^j \right).
    \end{split}
    \label{eq:HamiltonianGaussianRelativeModes}
\end{equation}
The dynamics generated by the latter Hamiltonian is illustrated in \autoref{fig:GaussianHamiltonianIllustration}. As a result of its simple form, we can provide a more detailed description of how the different terms in \eqref{eq:HamiltonianGaussianRelativeModes} contribute to the overall dynamics (see \autoref{fig:GaussianDistributionsIllustration}): The first term proportional to $\tilde{q}$ causes a local rotation (green dashed circle in \textbf{d)}), while the second term proportional to $\tilde{c}_1$ generates local squeezing (red dashed arrows in \textbf{d)}). Both act locally on the $j$-th. $+$ phase space. The coupling term proportional to $J$ builds up correlations between the wells (blue solid arrow in \textbf{i)}) by stretching the local distributions proportionally to each other (see \textbf{g)} versus \textbf{h)}).

One may also express \eqref{eq:HamiltonianGaussianRelativeModes} through the canonical field operators (see \hyperref[app:OperatorIdentitiesUndepletedPump]{Appendix} \ref{app:OperatorIdentitiesUndepletedPump} for details), leading to
\begin{equation}
    \begin{split}
        \boldsymbol{H}_{\text{up,Gauss}}^+ &= \frac{1}{2} \sum_{j=1}^N \Bigg[ \lambda^+ \left( \boldsymbol{\phi}_+^j \right)^2 + \lambda^- \left( \boldsymbol{\pi}_+^j \right)^2 \Bigg] \\
        &\hspace{0.5cm}-J \sum_{j = 1}^{N \text{ or } N-1} \left( \boldsymbol{\phi}_+^j \boldsymbol{\phi}_+^{j+1} +  \boldsymbol{\pi}_{+}^j \boldsymbol{\pi}_{+}^{j+1} \right) \\
        &\equiv\frac{1}{2} \sum_{j,j' = 1}^{2 N} \boldsymbol{\chi}^{j T}_{+} \, \Lambda^{j j'} \, \boldsymbol{\chi}^{j'}_{+},
    \end{split}
    \label{eq:HamiltonianGaussianCanonicalOperators}
\end{equation}
where we introduced the two new couplings
\begin{equation}
    \lambda^{\pm} = \tilde{q} \pm \tilde{c}_1
\end{equation}
for brevity\footnote{Note here that $\lambda^{\pm}$ does \textit{not} refer to the $\pm$ phase spaces.} and the so-called Hamiltonian matrix $\Lambda$. The latter is a $2N \times 2N$, real and symmetric matrix of block-diagonal form 
\begin{equation}
    \Lambda = \begin{pmatrix}
        \Lambda^+ & 0 \\
        0 & \Lambda^-
    \end{pmatrix},
\end{equation}
where $\Lambda^+$ and $\Lambda^-$ describe the dynamics of the fields $\boldsymbol{\phi}_+^j$ and the momentum fields $\boldsymbol{\pi}_+^j$, respectively. Both have the same form with non-vanishing entries only on the three leading diagonals for open boundary conditions, i.e.
\begin{equation}
    \Lambda^{\pm} = \begin{pmatrix}
    \lambda^{\pm} & -J & 0 & 0 \\
    -J & \lambda^{\pm} & -J & 0 \\
    0 & -J & \lambda^{\pm} & \ddots \\
    0 & 0 & \ddots & \ddots
    \end{pmatrix}.
    \label{eq:HamiltonianMatrices}
\end{equation}
Note that a contribution of $-J$ has to be added in the first and last anti-diagonal entries for periodic boundary conditions. Hence, the diagonal entries in \eqref{eq:HamiltonianMatrices} describe single wells, while the next-to-leading diagonals (and possibly the outer two entries of the anti-diagonal) contain the coupling between neighboring wells. We also note that the Hamiltonian \eqref{eq:HamiltonianGaussianCanonicalOperators} is bounded from below by a positive number, i.e., has a positive-energy ground state, when $\lambda^{\pm} \ge 0$, which is fulfilled if $q \ge c_1/2$ and hence in particular for non-negative $q \ge 0$.

\subsubsection{Mean and covariance matrix}
\label{subsubsec:MeanAndCovarianceMatrix}
It is well known that quadratic Hamiltonians map Gaussian states to Gaussian states \cite{Weedbrook2012,Serafini2017}. Since the initial state $\ket{0,\alpha,0}^{\otimes N}$ from Eq.~\eqref{eq:PolarState} corresponds to an uncorrelated set of local vacua in the $+$ phase spaces, the initial state in the undepleted pump approximation is indeed Gaussian. Hence, the dynamics generated by \eqref{eq:HamiltonianGaussianCanonicalOperators} is fully encoded in the two lowest-order moments of the quantum state $\boldsymbol{\rho}_+$, i.e., its field expectation values 
\begin{equation}
    \chi_{+}^j = \Tr \{\boldsymbol{\rho}_+ \, \boldsymbol{\chi}_{+}^j \}
    \label{eq:ExpectationValues}
\end{equation}
and its covariance matrix \cite{Weedbrook2012,Serafini2017}
\begin{equation}
    \gamma_{+}^{j j'} = \frac{1}{2} \, \Tr \{ \boldsymbol{\rho}_+ \{ \boldsymbol{\chi}_{+}^j - \chi_{+}^j, \boldsymbol{\chi}_{+}^{j'} - \chi_{+}^{j'} \} \}.
    \label{eq:CovarianceMatrix}
\end{equation}
The latter is a real, symmetric, and non-negative $2N$-dimensional matrix of block form
\begin{equation}
    \gamma_{+} = \begin{pmatrix}
    \mathcal{M}_{+} & \mathcal{T}_{+} \\
    \mathcal{T}_{+}^T & \mathcal{N}_{+}
    \end{pmatrix},
    \label{eq:CovarianceMatrix}
\end{equation}
and contains the two fundamental two-point correlation functions
\begin{equation}
    \begin{split}
        \mathcal{M}_{+}^{j j'} &= \Tr \left\{ \boldsymbol{\rho}_+ \, \boldsymbol{\phi}_{+}^j \boldsymbol{\phi}_{+}^{j'} \right\} - \phi_{+}^{j} \phi_{+}^{j'}, \\
        \mathcal{N}_{+}^{j j'} &= \Tr \left\{ \boldsymbol{\rho}_+ \, \boldsymbol{\pi}_{+}^j \boldsymbol{\pi}_{+}^{j'} \right\} - \pi_{+}^j \pi_{+}^{j'},
    \end{split}
    \label{eq:TwoPointCorrelators}
\end{equation}
as well as the mixed correlator
\begin{equation}
    \mathcal{T}_{+}^{j j'} = \Tr \left\{ \boldsymbol{\rho}_+ \, \boldsymbol{\phi}_{+}^j \boldsymbol{\pi}_{+}^{j'} \right\} - \frac{i}{2} \delta_{j j'} - \phi_{+}^j \pi_{+}^{j'}.
\end{equation}
Importantly, the covariance matrix is constrained by the Robertson-Schrödinger uncertainty relation in the form of a lower bound to its determinant \cite{Robertson1929,Robertson1930,Schroedinger1930}
\begin{equation}
    \det \gamma_+ \ge \frac{1}{2^{2N}},
\end{equation}
with equality if and only if $\boldsymbol{\rho}_+$ corresponds to a pure Gaussian state \cite{Weedbrook2012,Serafini2017}. Thus, for our analytical model, $\det \gamma_+ > 1/2^{2N}$ encodes mixedness of $\boldsymbol{\rho}_+$ (see also \autoref{subsubsec:SubtractedClassicalEntropies}). 

Importantly, $\chi_+$ and $\gamma_+$ correspond to the first- and second-order moments of the Wigner $W$-distribution $\mathcal{W}_+$, and the diagonal entries of $\gamma_+$ are encoded in $f_+$ and $g_+$. While $\chi_+$ is also the mean of the Husimi $Q$-distribution $\mathcal{Q}_+$, its covariance matrix $V_+$ acquires an additional contribution, which is a consequence of $\mathcal{Q}_+$ being a convolution of $\mathcal{W}_+$, resulting in
\begin{equation}
    V_+ = \gamma_+ + \frac{1}{2} \, \mathds{1}.
    \label{eq:CovarianceMatrixHusimiQ}
\end{equation}

\subsubsection{Symplectic time evolution}
\label{subsubsec:SymplecticTimeEvolution}
The effect of the evolution under the Hamiltonian \eqref{eq:HamiltonianGaussianCanonicalOperators} on the two lowest-order moments can be described by symplectic transformations in phase space \cite{Weedbrook2012,Serafini2017}. To that end, we recall a few basics of Gaussian quantum information theory. The Hamiltonian $\boldsymbol{H}^+_{\text{up,Gauss}}$ generates the time evolution of the initial state $\boldsymbol{\rho}_+ (0)$ at $t=0$ via the unitary time evolution operator
\begin{equation}
    \boldsymbol{U}^+_{\text{up,Gauss}}(t) = e^{-i t \boldsymbol{H}^+_{\text{up,Gauss}}},
    \label{eq:TimeEvolutionOperatorGaussian}
\end{equation}
in the sense that
\begin{equation}
    \boldsymbol{\rho}_+ (t) = \boldsymbol{U}^+_{\text{up,Gauss}} (t) \, \boldsymbol{\rho}_+ (0) \, \boldsymbol{U}^{+ \dagger}_{\text{up,Gauss}} (t).
\end{equation}
In phase space, the unitary transformation $\boldsymbol{U}^+_{\text{up,Gauss}}(t)$ corresponds to a symplectic transformation $\mathcal{S}^+_{\text{up,Gauss}}(t)$, that is, a transformation which preserves the commutator in phase space (see Eq. \eqref{eq:GroupedOperatorsCommutator}), i.e. leaves the symplectic form $\Omega$ invariant
\begin{equation}
    \mathcal{S}^{+}_{\text{up,Gauss}}(t) \, \Omega \, \mathcal{S}^{+,T}_{\text{up,Gauss}}(t) = \Omega.
\end{equation}
When the Hamiltonian is written as a bilinear form as in the second line in \eqref{eq:HamiltonianGaussianCanonicalOperators}, the symplectic time evolution matrix can be expressed through the Hamiltonian matrix $\Lambda$ via \cite{Weedbrook2012,Serafini2017}
\begin{equation}
    \mathcal{S}^+_{\text{up,Gauss}}(t) = e^{t \Omega \Lambda}.
\end{equation}
This matrix can directly be applied to the two lowest-order moments of $\boldsymbol{\rho}_+$, yielding the simple time evolution equations \cite{Weedbrook2012,Serafini2017}
\begin{equation}
    \begin{split}
        \chi_{+} (t) &= \mathcal{S}^+_{\text{up,Gauss}}(t) \, \chi_{+} (0), \\ 
        \gamma_+ (t) &= \mathcal{S}^+_{\text{up,Gauss}}(t) \, \gamma_+ (0) \, \mathcal{S}^{+T}_{\text{up,Gauss}}(t).
    \end{split}
    \label{eq:LowestMomentsTimeEvolution}
\end{equation}
Since the initial state is the uncoupled vacuum state, we have $\chi_+ (0) = 0$, and hence all phase-space distributions remain centered around the origin, i.e.
\begin{equation}
    \chi_{+} (t) = \chi_{+} (0) = 0.
\end{equation}
Instead, the initial covariance matrix 
\begin{equation}
    \gamma_+ (0) = \frac{1}{2} \, \mathds{1},
\end{equation}
evolves non-trivially following the second line of \eqref{eq:LowestMomentsTimeEvolution}, which can be computed analytically for a given set of parameters. 

Note that for a non-zero initial temperature, the initial covariance matrix is also proportional to the identity, but the variances increase due to thermal fluctuations, which results in
\begin{equation}
    \gamma_+ (0) = \left[ \frac{1}{2} + n_{\text{BE}} (\beta) \right] \mathds{1}.
    \label{eq:ThermalAnalytics}
\end{equation}
Note also that adding half the identity to $\gamma_+ (t)$ gives the time-evolved covariance matrix $V_+ (t)$ of the Husimi $Q$-distribution. Since $\mathcal{S} \mathcal{S}^T = \mathds{1}$ for all symplectic matrices $\mathcal{S}$ one may equally apply the symplectic transformation directly to the defining equation of $V_+$, i.e. \eqref{eq:CovarianceMatrixHusimiQ}.

\subsubsection{Comment on field theory correspondence}
\label{subsubsec:FieldTheoryCorrespondence}
At last, we comment on the field theory the Gaussian Hamiltonian \eqref{eq:HamiltonianGaussianCanonicalOperators} reproduces when taking the continuum limit. To that end, we introduce a lattice spacing $\epsilon$ and define the field operators
\begin{equation}
    \boldsymbol{\phi}_+ (x) \equiv \frac{1}{\sqrt{\epsilon}} \, \boldsymbol{\phi}_+^j, \quad \boldsymbol{\pi}_+ (x) \equiv \frac{1}{\sqrt{\epsilon}} \, \boldsymbol{\pi}_+^j,
    \label{eq:CanonicalOperatorsContinuum}
\end{equation}
which fulfill the distribution-valued commutation relations
\begin{equation}
    [\boldsymbol{\phi}_+ (x), \boldsymbol{\pi}_+ (x') ] = i \delta (x - x'),
\end{equation}
in the continuum limit $\epsilon \to 0$. Then, using \eqref{eq:CanonicalOperatorsContinuum} in \eqref{eq:HamiltonianGaussianCanonicalOperators} together with
\begin{equation}
    \boldsymbol{\phi}_+ (x + \epsilon) = \boldsymbol{\phi}_+ (x) + \epsilon \,  \partial_x \boldsymbol{\phi}_+ (x) + \mathcal{O}(\epsilon^2), 
\end{equation}
analogously for $\boldsymbol{\pi}_+ (x + \epsilon)$, results in
\begin{equation}
    \boldsymbol{H}^+_{\text{up,Gauss}} = \frac{1}{2} \int \mathrm{d} x \, \left[ \kappa^+ (x) \, \boldsymbol{\phi}_+^2 (x) + \kappa^- (x) \, \boldsymbol{\pi}_+^2 (x) \right],
\end{equation}
with the differential operators
\begin{equation}
    \kappa^{\pm} (x) = \lambda^{\pm} + 1 + J \epsilon \, \partial_x,
\end{equation}
to leading order in $\epsilon$. Hence, a reasonable continuum limit requires the limit of infinite coupling $J \to \infty$ with the product $J \epsilon$ kept fixed, which effectively implements $\epsilon = 1/J \to 0$.

\section{Area law in phase space}
\label{sec:AreaLaw}
We now discuss the scaling behaviors of quantum entropies associated with a subsystem. Then, we define \textit{classical} entropies of the phase space distributions introduced in \autoref{subsec:PhaseSpaceDistributions} and provide mathematical as well as heuristic arguments for why these entropies encode the area law.

\subsection{Quantum entropies}
\label{subsec:QuantumEntropies}
\begin{figure}[t!]
    \centering
    \includegraphics[width=0.9\columnwidth]{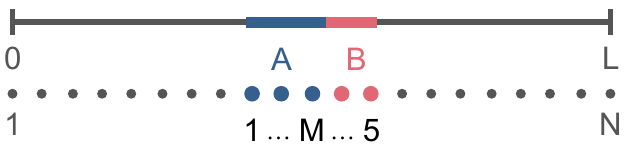}
    \caption{The open subsystem of interest (colored) is partitioned into $A$ (blue) and $B$ (red).}
    \label{fig:SpatialSubregion}
\end{figure}

We consider a subsystem of five wells from the $N$ total wells, which can exchange energy and particles with its complement and, therefore, should be considered an open quantum system. We partition this subsystem into a subregion $A$ consisting of the first $0 \le M \le 5$ wells and its complement $B$ composed out of the remaining $5-M$ wells (see \autoref{fig:SpatialSubregion}). The multi-well setup offers the possibility to study the discretized version of a continuous quantum field theory with the wells corresponding to lattice points. The local states associated with the subregions are, as usual, defined via the partial trace
\begin{equation}
    \boldsymbol{\rho}_+^{A} = \Tr_{B} \{ \boldsymbol{\rho}_+ \}
\end{equation}
and analogous for $\boldsymbol{\rho}_+^{B}$.
The local mixedness of $A$ is conveniently measured by the von Neumann entanglement entropy \cite{vonNeumann1955}
\begin{equation}
    S (\boldsymbol{\rho}_+^A) = - \Tr \{ \boldsymbol{\rho}_+^A \, \ln \boldsymbol{\rho}_+^A \},
    \label{eq:EntanglementEntropy}
\end{equation}
which serves as an entanglement measure if and only if the bipartite state $\boldsymbol{\rho}_+$ is pure, in which case $S (\boldsymbol{\rho}_+^A) > 0$ if and only if $A$ and $B$ are entangled and $S (\boldsymbol{\rho}_+^A) = 0$ if not. However, when the bipartite state is mixed, for example, due to the presence of thermal fluctuations or coupling to the environment as considered here, one typically studies the quantum mutual information instead. It is defined as \cite{Nielsen2010,Wilde2013}
\begin{equation}
    I (\boldsymbol{\rho}_+^A : \boldsymbol{\rho}_+^B) = \Tr \left\{ \boldsymbol{\rho}_+ \left( \ln \boldsymbol{\rho}_+ - \ln \boldsymbol{\rho}_+^A \otimes \boldsymbol{\rho}_+^B \right) \right\},
    \label{eq:QuantumMutualInformation}
\end{equation}
or equivalently as
\begin{equation}
    I (\boldsymbol{\rho}_+^A : \boldsymbol{\rho}_+^B) = S (\boldsymbol{\rho}^A_+) + S (\boldsymbol{\rho}^B_+) - S (\boldsymbol{\rho}_+),
    \label{eq:QuantumMutualInformation2}
\end{equation}
if all three terms on the right are finite, and serves as the measure for the total correlations between $A$ and $B$ being zero if and only if $\boldsymbol{\rho}_+ = \boldsymbol{\rho}_+^A \otimes \boldsymbol{\rho}_+^B$. Note that when the initial state is pure, the entanglement entropy and the quantum mutual information are proportional to each other, i.e., $I (\boldsymbol{\rho}_+^A : \boldsymbol{\rho}_+^B) = 2 S (\boldsymbol{\rho}_+^A)$.

Both quantities have been studied extensively for a large variety of quantum field theories and quantum many-body systems, and it is well-known that for \textit{typical} states of Hamiltonians with local interactions, they scale with the \textit{area} of the entangling surface separating $A$ and $B$, see e.g. \cite{Calabrese2004,Calabrese2006,Calabrese2009,Casini2009,Eisert2010} for reviews. In this context, states are called typical, for example, when they lie sufficiently close to the ground state of the Hamiltonian or when they are generated from an initial product state on short time scales after a quench \cite{Eisert2010}. In the following, we shall be concerned with the latter scenario. 

\begin{figure}[t!]
    \centering
    \includegraphics[width=0.95\columnwidth]{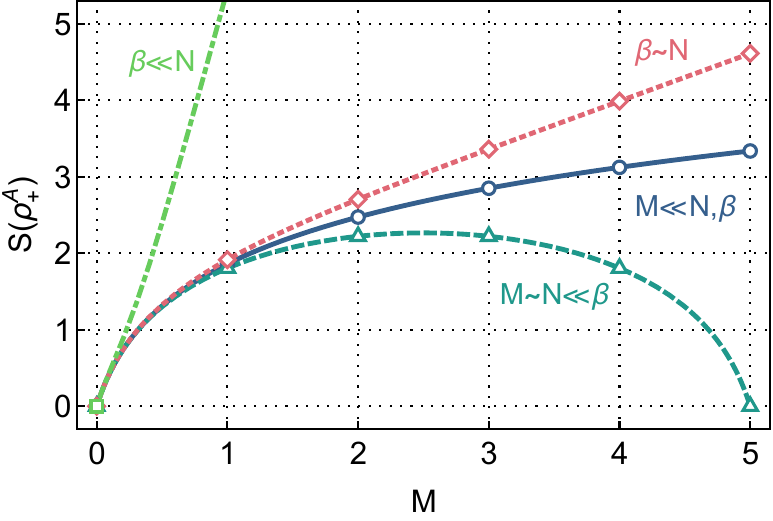}
    \caption{Scaling of the entanglement entropy $S(\boldsymbol{\rho}_+^A)$ with subsystem size $M$ for the four archetypal scenarios: large system size at zero temperature (blue curve), finite system size at zero temperature (petrol curve), finite temperature (red curve) and high temperature (green curve). For small $M$, the first three curves converge.}
    \label{fig:EntropyScaling}
\end{figure}

For our setup, three competing length scales are relevant for the precise scaling with subsystem size $M$ of the aforementioned quantum entropies: the system size $N$, the (initial) inverse temperature $\beta = 1/T$, and the subsystem size $M$ itself. When the dominating scale is $M$, i.e. when we consider the subinterval size being small compared to the system size and the inverse temperature $M \ll N, \beta$, then the entanglement entropy scales logarithmically (blue solid curve in \autoref{fig:EntropyScaling}) \cite{Calabrese2004,Calabrese2006,Calabrese2009,Casini2009}
\begin{equation}
    S (\boldsymbol{\rho}_+^A) \sim \kappa_1 \ln \left( M + \kappa_2 \right) + \kappa_3,
    \label{eq:AreaLaw}
\end{equation}
with $\kappa_i$ being real constants. We normalize all entropic quantities such that they vanish at $M=0$, resulting in the constraint
\begin{equation}
    \kappa_2 = e^{- \kappa_3 / \kappa_1},
    \label{eq:AreaLawConstraint}
\end{equation}
and hence only $\kappa_1$ and $\kappa_3$ are independent. Typically, $\kappa_1$ is universal, i.e., regularization independent, especially in conformal field theories where it corresponds to the central charge, while $\kappa_3$ is usually not \cite{Calabrese2004, Calabrese2006, Calabrese2009}.

When instead $N$ is dominant, i.e., when the subinterval size is of the order of the system size $M \sim N$ and the temperature is still small $N \ll \beta$, the so-called finite-size area law holds (petrol dashed curve) \cite{Calabrese2004,Calabrese2006}
\begin{equation}
    S (\boldsymbol{\rho}_+^A) \sim \kappa_1 \ln \left[ \frac{N}{\pi} \sin \left(\frac{\pi M}{N} \right) + \kappa_2 \right] + \kappa_3,
    \label{eq:FiniteSizeAreaLaw}
\end{equation}
again subject to the constraint \eqref{eq:AreaLawConstraint}. Since the quantum state of the whole system is close to being pure in this case, we have $S (\boldsymbol{\rho}_+^A) \to 0$ when $M \to N$. Further, in the limit of small $M \ll N$, the latter equation reduces to \eqref{eq:AreaLaw} since $(N/\pi) \sin (\pi M/N) = M$ to first order in $M$.

If finite-temperature effects become relevant, i.e., for $\beta \sim N$, we instead expect a scaling of the form (red dotted curve) \cite{Calabrese2004,Calabrese2006}
\begin{equation}
    S (\boldsymbol{\rho}_+^A) \sim \kappa_1 \ln \left[ \frac{\beta}{\pi} \sinh \left( \frac{ \pi M}{\beta} \right) + \kappa_2 \right] + \kappa_3,
    \label{eq:FiniteTemperatureAreaLaw}
\end{equation}
again constrained by \eqref{eq:AreaLawConstraint}. In this case, the entropy remains finite if computed for the whole system due to additional (classical) mixedness. While for small temperatures $\beta \gg N$ we obtain back \eqref{eq:AreaLaw}, large temperatures $\beta \ll N$ lead to an extensive entropy obeying a volume law (green dot-dashed curve)
\begin{equation}
    S (\boldsymbol{\rho}_+^A) \sim \kappa_1 M.
    \label{eq:VolumeLaw}
\end{equation}
In this case, classical correlations dominate over quantum ones. 

Remarkably, the quantum mutual information follows the finite-size area law \eqref{eq:FiniteSizeAreaLaw} in all of the four considered cases \cite{Wolf2008}. In particular, even if thermal fluctuations prevent the appearance of the area law in the entropies, the extensive contributions drop out when considering the quantum mutual information, which demonstrates its utility for observing an area law in actual experiments.

\subsection{Classical entropies}
\label{subsec:ClassicalEntropies}
We now introduce various entropic measures in the phase space associated with subsystem $A$ (subsystem $B$ and the full subsystem $AB$ can be treated on equal footing). To that end, we shall group the set of corresponding local field variables according to
\begin{equation}
    \chi_+^A = (\phi_+^A, \pi_+^A)^T = (\phi_+^1, ..., \phi_+^M, \pi_+^1, ..., \pi_+^M)^T.
\end{equation}
Then, the distributions corresponding to $A$ are obtained by integrating out all degrees of freedom belonging to subsystem $B$, for instance
\begin{equation}
    \mathcal{W}_+^A = \int \mathrm{d} \phi^B_{+} \, \mathrm{d} \pi_+^B \, \mathcal{W}_+,
\end{equation}
and similarly for the other two types of phase space distributions.

\subsubsection{Standard entropies}
Let us start with the Wigner entropy, which is defined as the differential entropy of the Wigner $W$-distribution, namely
\begin{equation}
    S (\mathcal{W}_+^A) = - \int \mathrm{d} \phi^A_{+} \, \mathrm{d} \pi_+^A \, \mathcal{W}_+^A \, \ln \mathcal{W}_+^A.
    \label{eq:WignerEntropy}
\end{equation}
It is well-defined, i.e., real, provided that $\mathcal{W}_+^A \ge 0$, otherwise it becomes complex-valued. A lower bound encoding the uncertainty principle has been conjectured in \cite{VanHerstraeten2021a,VanHerstraeten2023,VanHerstraeten2021b} and reads
\begin{equation}
    S (\mathcal{W}_+^A) \ge S (\bar{\mathcal{W}}_+^A) =  M (1 + \ln \pi),
    \label{eq:WignerEntropyConjecture}
\end{equation}
with equality if and only if $\mathcal{W}_+^A$ corresponds to a product of pure Gaussian states.

The marginal entropy of $f_+^A$ is defined as
\begin{equation}
    S (f_+^A) = - \int \mathrm{d} \phi^A_{+} \, f_+^A \, \ln f_+^A,
    \label{eq:MarginalEntropy}
\end{equation}
and analogously for $g_+^A$. Since $f_+^A$ and $g_+^A$ are true probability density functions, their entropies are always well-defined. The corresponding entropic uncertainty relation has been put forward by Bia{\l}ynicki-Birula and Mycielski \cite{Everett1957,Hirschman1957,Beckner1975,Bialynicki-Birula1975} (see also \cite{Coles2017,Hertz2019} for reviews)
\begin{equation}
    S (f_+^A) + S (g_+^A) \ge S (\bar{f}_+^A) + S (\bar{g}_+^A) = M (1 + \ln \pi),
    \label{eq:BBMEUR}
\end{equation}
which contains the same bound as the Wigner entropy conjecture \eqref{eq:WignerEntropyConjecture}. In fact, \eqref{eq:BBMEUR} would be a simple consequence of \eqref{eq:WignerEntropyConjecture} when using the subadditivity of entropy, namely
\begin{equation}
    S (\mathcal{W}_+^A) \le S (f_+^A) + S (g_+^A),
\end{equation}
with equality if and only if $W^A_{+} = f^{A}_{+} \, g^A_{+}$. However, the bound in \eqref{eq:BBMEUR} is less tight since it is attained only for products of squeezed vacuum states.

At last, we introduce the Wehrl entropy \cite{Wehrl1978,Wehrl1979}
\begin{equation}
    S (\mathcal{Q}_+^A) = - \int \frac{\mathrm{d}\phi^A_+ \, \mathrm{d}\pi^A_+}{(2 \pi)^M} \, \mathcal{Q}_+^A \ln \mathcal{Q}_+^A,
    \label{eq:WehrlEntropy}
\end{equation}
which is also always well-defined since $\mathcal{Q}_+^A \ge 0$. It is bounded from below by the Wehrl-Lieb inequality \cite{Wehrl1978,Wehrl1979,Lieb1978,Lieb2014}
\begin{equation}
    S (\mathcal{Q}_+^{A}) \ge S (\bar{\mathcal{Q}}_+^A) = M,
    \label{eq:WehrlLiebInequality}
\end{equation}
which is tight if and only if the state is a product of pure coherent states. It is also bounded by the Wigner and quantum entropies \cite{Haas2021b}
\begin{equation}
     S (\mathcal{Q}_+^{A}) \ge S (\mathcal{W}_+^{A}) - M \ln \pi, \quad S (\mathcal{Q}_+^{A}) \ge S (\boldsymbol{\rho}_+^A).
\end{equation}
Note here that the additional constant $-M \ln \pi$ is a result of the different normalizations of the Wigner $W$- and Husimi $Q$-distributions. Therefore, \eqref{eq:WehrlLiebInequality} would also be a direct consequence of \eqref{eq:WignerEntropyConjecture}.

The three aforementioned types of entropies have in common that their lower bounds \eqref{eq:WignerEntropyConjecture}, \eqref{eq:BBMEUR} and \eqref{eq:WehrlLiebInequality} are attained for a product of pure vacuum states and scale with the volume of the subsystem $A$. In this sense, all these entropies are \textit{classical} to leading order for arbitrary states with the main contribution coming from the vacuum, in contrast to the quantum entropies introduced in \autoref{subsec:QuantumEntropies}, which vanish for all pure states. In what follows, we will argue that quantum features such as the area law are still present in classical entropies but hidden in the next-to-leading order terms.

\subsubsection{Subtracted entropies}
\label{subsubsec:SubtractedClassicalEntropies}
Recently, it has been argued that quantum features such as the area law of the entanglement entropy are also present in classical entropies when the leading order extensive contributions are subtracted \cite{Haas2023b}. Based on this idea, we define \textit{subtracted} classical entropies by subtracting the extensive vacuum contributions appearing in the bounds of the corresponding entropic uncertainty relations, to wit
\begin{equation}
    \begin{split}
        \Delta S (\mathcal{W}_+^A) &\equiv S (\mathcal{W}_+^A) - S (\bar{\mathcal{W}}_+^A), \\
        \Delta S (f_+^A,g_+^A) &\equiv S (f_+^A) + S (g_+^A) - S (\bar{f}_+^A) - S (\bar{g}_+^A), \\
        \Delta S (\mathcal{Q}_+^A) &\equiv S (\mathcal{Q}_+^A) - S (\bar{\mathcal{Q}}_+^A).
    \end{split}
    \label{eq:SubtractedClassicalEntropies}
\end{equation}
Let us provide some simple arguments for why these entropies encode the very same features as their quantum analogs in the following. 

First, we consider Gaussian quantum states, which provide a reasonable first-order approximation to the states generated by the Hamiltonian \eqref{eq:FullHamiltonian}, see \autoref{subsec:GaussianModel}. A general Gaussian Wigner $W$-distribution is of the form
\begin{equation}
    \mathcal{W}_+^A = \frac{1}{Z^{A}_+} \, e^{- \frac{1}{2} (\chi_+^{A})^T (\gamma_+^A)^{-1} \chi_+^A},
    \label{eq:GaussianWignerW}
\end{equation}
with $Z^{A}_+ = (2\pi)^M \, \sqrt{\det \gamma_+^A}$ being a normalization constant. Computing the subtracted Wigner entropy for such a state gives \cite{Adesso2012,Serafini2017}
\begin{equation}
    \Delta S (\mathcal{W}_+^{A}) = \frac{1}{2} \ln \left(2^{2 M} \det \gamma_+^A \right) = S_2 (\boldsymbol{\rho}_+^A),
    \label{eq:SubtractedWignerEntropyGaussian}
\end{equation}
where
\begin{equation}
    S_2 (\boldsymbol{\rho}_+^{A}) = - \ln \Tr \{(\boldsymbol{\rho}_+^{A})^2 \}
    \label{eq:Renyi2Entropy}
\end{equation}
denotes the quantum Rényi-2 entropy of the state $\boldsymbol{\rho}^A_+$ that corresponds to $\mathcal{W}^A_+$. Hence, the subtracted Wigner entropy \textit{coincides} with a quantum entropy for Gaussian states. For completeness, let us also give the Gaussian expressions for the Husimi-based quantities, which read
\begin{equation}
    \mathcal{Q}_+^A = \frac{1}{Z^{A}_+} \, e^{- \frac{1}{2} (\chi_+^{A})^T (V_+^A)^{-1} \chi_+^A},
    \label{eq:GaussianHusimiQ}
\end{equation}
with $Z^{A}_+ = \sqrt{\det V_+^A}$ and 
\begin{equation}
    \Delta S (\mathcal{Q}_+^{A}) = \frac{1}{2} \ln \det V_+^A.
    \label{eq:SubtractedWehrlEntropyGaussian}
\end{equation}
Note, however, that these entropies can not be related to quantum entropies in general.

For arbitrary states, including Wigner-negative states, a general relation between subtracted classical and quantum entropies can only be established for the subtracted Rényi-2 Wigner entropy \cite{Manfredi2000,Wlodarz2003}
\begin{equation}
    \Delta S_2 (\mathcal{W}_+^A) = S_2 (\boldsymbol{\rho}_+^A).
    \label{eq:Renyi2WignerEntropy}
\end{equation}
Note here that the entropic order alters the lower bound in the corresponding entropic uncertainty relation, and hence a different term compared to $\Delta S (\mathcal{W}_+^A)$ has to be subtracted. Although the relation \eqref{eq:Renyi2WignerEntropy} can \textit{not} be generalized to other subtracted Rényi-Wigner entropies, it has been shown in \cite{Haas2023b} that the crucial feature of quantum entropies, i.e. the area law, is present for the entire family of subtracted Rényi-Wigner entropies beyond the Gaussian case, indicating that the area law may also appear for the classical entropies defined in \eqref{eq:SubtractedClassicalEntropies}, which we investigate in detail in \autoref{sec:Results}.

\subsection{Classical mutual informations}
\label{subsec:ClassicalMutualInformations}
Let us also introduce classical mutual informations as measures for correlations in phase space. For Wigner-positive states, we define the Wigner mutual information
\begin{equation}
    I (\mathcal{W}_+^A : \mathcal{W}_+^B) = S(\mathcal{W}_+^A) + S(\mathcal{W}_+^B) - S(\mathcal{W}_+),
    \label{eq:WignerMutualInformation}
\end{equation}
while for general states, we define the marginal mutual informations
\begin{equation}
    I (f_+^A : f_+^B) = S(f_+^A) + S(f_+^B) - S(f_+),
    \label{eq:MarginalMutualInformation}
\end{equation}
(similarly for $g_+$) and the Wehrl mutual information \cite{Haas2021b} 
\begin{equation}
    I (\mathcal{Q}_+^A : \mathcal{Q}_+^B) = S(\mathcal{Q}_+^A) + S(\mathcal{Q}_+^B) - S(\mathcal{Q}_+).
    \label{eq:WehrlMutualInformation}
\end{equation}
All of them are non-negative functionals, being zero if and only if the two local distributions on $A$ and $B$ are uncorrelated. 

It has been shown that every classical mutual information is a lower bound to the quantum mutual information \cite{Lieb2005,Haas2021b}. In particular, we have
\begin{equation}
    I (\mathcal{W}_+^A : \mathcal{W}_+^B), \, I (\mathcal{Q}_+^A : \mathcal{Q}_+^B) \le I (\boldsymbol{\rho}_+^A : \boldsymbol{\rho}_+^B),
    \label{eq:ClassicalMutualInformationBound}
\end{equation}
indicating that classical mutual informations do \textit{not} capture all correlations in general. However, the bound \eqref{eq:ClassicalMutualInformationBound} is expected to be tighter than the standard second-order lower bounds on the quantum mutual information in terms of two-point correlation functions (which only faithfully describe Gaussian correlations).

In contrast to classical entropies, extensive contributions naturally cancel when considering classical mutual informations. For example, decomposing \eqref{eq:WignerMutualInformation} in the sense of \eqref{eq:QuantumMutualInformation2} shows that
\begin{equation}
    I (\mathcal{W}_+^A : \mathcal{W}_+^B) = \Delta S (\mathcal{W}_+^A) + \Delta S (\mathcal{W}_+^B) - \Delta S (\mathcal{W}_+),
\end{equation}
since the vacuum is uncorrelated, i.e. $S (\bar{\mathcal{W}}_+) = S (\bar{\mathcal{W}}_+^A) + S (\bar{\mathcal{W}}_+^B)$. For Gaussian states, we can find a relation analogous to \eqref{eq:SubtractedWignerEntropyGaussian}, which reads \cite{Adesso2012}
\begin{equation}
    \begin{split}
        I (\mathcal{W}_+^A : \mathcal{W}_+^B) &= \frac{1}{2} \ln \frac{\det \gamma_+^A \, \det \gamma_+^B}{\det \gamma_+} \\
        &= I_2 (\boldsymbol{\rho}_+^{A} : \boldsymbol{\rho}_+^{B}),
    \end{split}
    \label{eq:WignerMutualInformationGaussian}
\end{equation}
with the Rényi-2 mutual information being defined as
\begin{equation}
    I_2 (\boldsymbol{\rho}_+^{A} : \boldsymbol{\rho}_+^{B}) = S_2 (\boldsymbol{\rho}_+^{A}) + S_2 (\boldsymbol{\rho}_+^{B}) - S_2 (\boldsymbol{\rho}_+).
    \label{eq:Renyi2MutualInformation}
\end{equation}
Fortunately, the partly heuristic arguments for the appearance of the area law for classical entropies in \autoref{subsubsec:SubtractedClassicalEntropies} can be made rigorous when considering classical mutual informations by adapting the argument in \cite{Wolf2008}. To that end, we first note that all three Hamiltonians \eqref{eq:FullHamiltonian}, \eqref{eq:HamiltonianUndepletedPumpDecomposition} and \eqref{eq:HamiltonianGaussianRelativeModes} can be decomposed according to
\begin{equation}
    \boldsymbol{H} = \boldsymbol{H}^A + \boldsymbol{H}^B + \boldsymbol{H}^{\partial},
    \label{eq:BipartiteHamiltoniansDecomposition}
\end{equation}
where $\boldsymbol{H}^A$ and $ \boldsymbol{H}^B$ denote the Hamiltonians of subsystems $A$ and $B$, respectively, while $\boldsymbol{H}^{\partial}$ contains the local interactions coupling $A$ to $B$. Further, we consider the rather general class of thermal states for a given Hamiltonian $\boldsymbol{H}$, i.e.
\begin{equation}
    \boldsymbol{\rho}_+ = \frac{1}{Z} \, e^{- \beta \boldsymbol{H}},
    \label{eq:ThermalState}
\end{equation}
which are typically of non-Gaussian form. Then, every distribution $\mathcal{O}_+$ over measurement outcomes $o_+$ corresponding to some POVM $\boldsymbol{O}_+$ can be written as
\begin{equation}
    \mathcal{O}_+ = \frac{1}{Z} \, e^{- \beta H},
\end{equation}
with the classical Hamiltonian $H$ being defined implicitly via
\begin{equation}
    \Tr \{e^{- \boldsymbol{H}} \, \boldsymbol{O}_+ \} = e^{- H}.
\end{equation}
For such a distribution, the classical entropy evaluates to
\begin{equation}
    S (\mathcal{O}_+) = \ln Z + \beta \braket{H}_{\mathcal{O}_+},
    \label{eq:ClassicalEntropyThermal}
\end{equation}
where
\begin{equation}
    \braket{H}_{\mathcal{O}_+} = \int \mathrm{d}o_+ \, \mathcal{O}_+ (o_+) \, H
\end{equation}
is the classical energy expectation value with respect to the distribution $\mathcal{O_+}(o_+)$. Using \eqref{eq:ClassicalEntropyThermal}, we find that the (classical) free energy is nothing but
\begin{equation}
    F (\mathcal{O}_+) \equiv - \frac{\ln Z}{\beta} = \braket{H}_{\mathcal{O}_+} - \frac{S (\mathcal{O}_+)}{\beta}.
    \label{eq:ClassicalFreeEnergyThermal}
\end{equation}
The latter is minimized by the thermal distribution $\mathcal{O}_+$, a fact which can be derived, for instance, using the non-negativity of the classical relative entropy of any given distribution with respect to the thermal distribution $\mathcal{O}_+$. As a special case, we can conclude that the thermal free energy is bounded from above by the free energy of the corresponding product distribution, i.e.
\begin{equation}
    F (\mathcal{O}_+) \le F (\mathcal{O}_+^A \, \mathcal{O}_+^B).
\end{equation}
Expanding the latter inequality in the sense of \eqref{eq:ClassicalFreeEnergyThermal} and using the additivity of classical entropies for product distributions yields
\begin{equation}
    \begin{split}
        &S (\mathcal{O}_+^A) + S( \mathcal{O}_+^B) - S (\mathcal{O}_+) \\
        &\le \beta \left(\braket{H}_{\mathcal{O}^A_+ \, \mathcal{O}^B_+} -\braket{H}_{\mathcal{O}_+} \right).
    \end{split}
\end{equation}
The decomposition \eqref{eq:BipartiteHamiltoniansDecomposition} implies
\begin{equation}
    \begin{split}
        \braket{H}_{\mathcal{O}^A_+ \, \mathcal{O}^B_+} -\braket{H}_{\mathcal{O}_+} &= \braket{H^{\partial}}_{\mathcal{O}^A_+ \, \mathcal{O}^B_+} - \braket{H^{\partial}}_{\mathcal{O}_+},
    \end{split}
\end{equation}
which, together with the definition of the classical mutual information, finally leads to
\begin{equation}
    I (\mathcal{O}^A_+ : \mathcal{O}^B_+) \le \beta \left( \braket{H^{\partial}}_{\mathcal{O}^A_+ \, \mathcal{O}^B_+} - \braket{H^{\partial}}_{\mathcal{O}_+} \right).
\end{equation}
The derived upper bound on the classical mutual information is a function of the classical boundary Hamiltonian $H^{\partial}$ only. This proves the area law for all variants of classical mutual information, including the three of our interest, for local interactions and thermal states.

\section{Entropy Estimation}
\label{sec:EntropyEstimation}
We now turn to the challenging task of estimating classical entropies from finitely many samples. Although this is a general problem, it arises especially for the system of our interest since experimental runs are particularly costly. To that end, we rely on suitable $k$-nearest-neighbor estimators and benchmark their validity for various scenarios. For our implementation, we utilize the `Non-parametric Entropy Estimation Toolbox' for Python, publicly available at \cite{SteegGitHub}.

\begin{figure*}
    \centering
    \includegraphics[width=.9\linewidth]{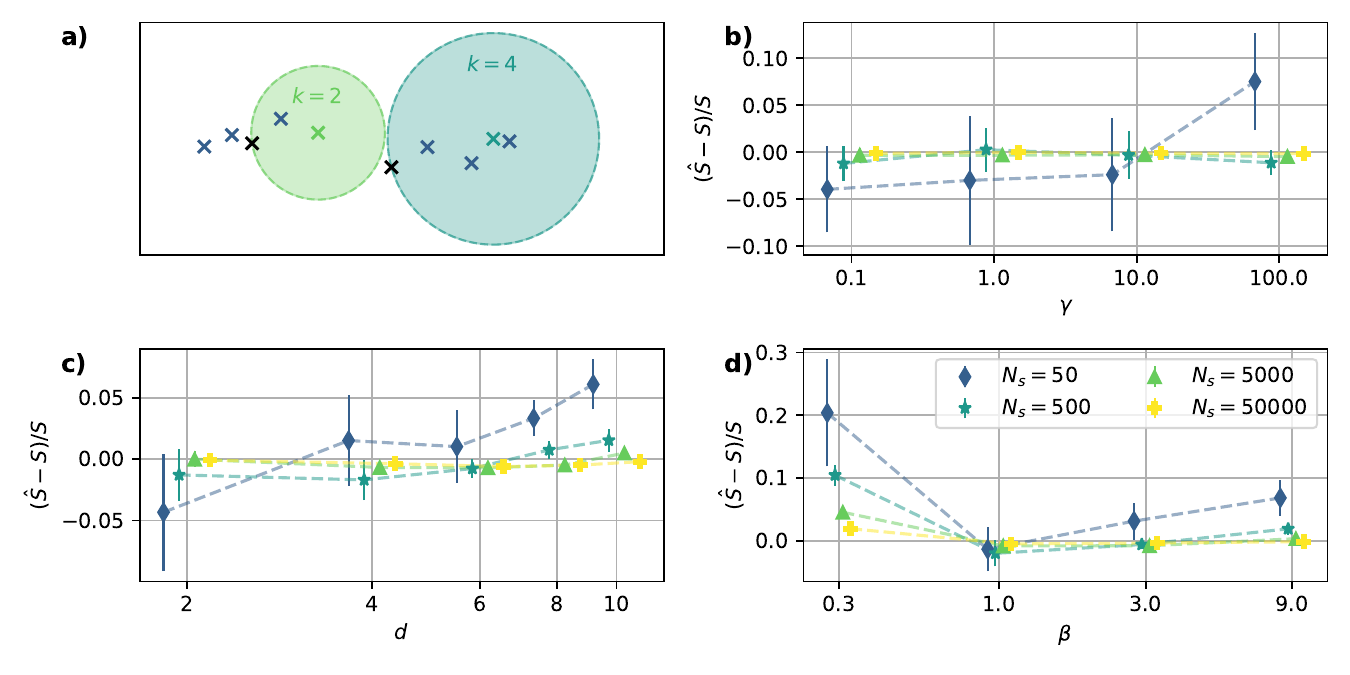}
    \caption{Benchmark of the entropy estimator $\hat{S}$ for distributions that allow for an analytic comparison to $S$. In \textbf{a)}, we visualize the algorithm for a few Gaussian samples in $d=2$ dimensions, with two random samples selected and two choices for the hyperparameter $k$, namely $k=2$ and $k=4$. The only assumption of the estimator is constant density in the light green and cyan circles, respectively, an assumption less justified by the larger $k$. Both $k$-th neighbors are shown in black. In \textbf{b)}, we benchmark the entropy estimator on a squeezed covariance matrix in $d=2$ dimensions with squeezing parameter $\gamma$, where large $\gamma$ correspond to strong squeezing. The covariance matrix is given by $\Sigma=(\mathds{1} + \gamma J) / \det(\mathds{1} + \gamma J)^{1/d}$ where $J$ is the matrix of all ones. All error bars are obtained from ten independent estimates, and we use offsets on the $x$ axis to differentiate the different data points. In \textbf{c)}, we let $\gamma=0$ and vary the dimension $d$, observing decreasing performance with increasing dimension, as is expected. In \textbf{d)}, we tune away from the Gaussian regime in $d=4$ dimensions, using a generalized normal distribution \cite{Nadarajah2005} parameterized by $\beta$ with scale $\alpha$ chosen such that the entropy $S$ takes the value of a standard normal distribution in $d=4$ dimensions. As can be seen from the density function in Eq.~\eqref{eq:GeneralizedNormalPDF}, small values of $\beta$ correspond to heavy tails, $\beta=2$ corresponds to a Gaussian shape, and the support of the distribution shrinks to the interval $[-1, 1]$ for $\beta\to \infty$. We denote the sample set size with $N_s$ and choose the nearest neighbor parameter $k$ to be three in all experiments.}
    \label{fig:EntropyEstimatorBenchmark}
\end{figure*}

\subsection{$k$-nearest-neighbor estimator}
\label{subsec:kNNEstimator}
The integrals that give the Wigner or Wehrl entropy, Eqs.~\eqref{eq:WignerEntropy} and \eqref{eq:WehrlEntropy}, respectively, are in general hard to estimate from samples. Unless the distribution is of a specific form, such as a Gaussian, that allows reexpressing the entropy through low-order correlators, one must use the available samples to approximate local densities of the distribution. A straightforward approach is to bin the samples onto a grid and use the relative frequencies in the histogram as approximations for the local densities. While this procedure presents an asymptotically unbiased estimator in the limit of infinite samples and small bins, it is problematic in practice since the binning procedure presents an information bottleneck as it coarse grains the information about the precise sample position. 

A more elaborate approach is to use the $k$-nearest-neighbor ($k$NN) statistics of the sample set to approximate local densities \cite{Kozachenko1987,Gyrfi1987,Joe1989, Hall1993,Birge1995,Beirlant1997,Singh2003,Kraskov2004,Goria2005,Sricharan2013,Kandasamy2015,Singh2016,Berrett2016,Gao2018,Lu2020}. The intuition behind the nearest-neighbor statistics is that samples with short distances to their neighbors are located in regions of high probability density, while samples which are located in regions of low probability density have nearest neighbors that are far away. A formalization of this intuition allows to construct an asymptotically unbiased discretization-free estimator $\hat{S}$ for differential entropies \cite{Kozachenko1987, Kraskov2004}
\begin{equation}
    \hat{S}(k, N_{\text{s}}) = g(k, N_{\text{s}}, d) + \frac{d}{N_{\text{s}}} \sum_{i=1}^{N_{\text{s}}} \ln \epsilon^i (k)
    \label{eq:EntropyEstimator}
\end{equation}
where $d$ is the dimension of the underlying distribution (for $M$ modes, we have $d=N$ for marginal and $d=2N$ for phase space distributions), $N_{\text{s}}$ is the number of samples and $\epsilon^i(k)$ is the distance of the $i$-th sample to its $k$-th neighbor. The term $g(k, N_{\text{s}}, d)$ is a constant that is independent of the samples and is given by \cite{Kozachenko1987, Kraskov2004}
\begin{equation}
    g(k, N_{\text{s}}, d) = -\psi(k) + \psi(N_{\text{s}}) + \ln(c_d),
    \label{eq:EntropyEstimatorSysError}
\end{equation}
where $\psi$ is the digamma function and $c_d$ is the volume of the $d$-dimensional unit ball. The parameter $k \in \mathbb{N}$ is a free parameter that should be chosen such that statistical and systematic errors are balanced. For close-to-Gaussian distributions in low dimensions, i.e., $d \lesssim 10$, $k=3$ is a convenient choice \cite{Kraskov2004,SteegGitHub}, which we adapt throughout this work.

The asymptotic unbiasedness of the estimator means that one is guaranteed to obtain the true entropy in the limit of infinitely many samples, which directly stems from the observation that the sample distances must shrink to zero in this limit. This is in contrast to approaches that fit a probability density function of a specific form to the data, which may result in more accurate estimates in the regime of few samples but will generally not converge to the true value when the amount of samples is increased.

We aim to give an intuitive picture of the $k$NN method in \autoref{fig:EntropyEstimatorBenchmark} \textbf{a)}, where the spheres, for which constant density is assumed, are drawn for $k=2$ and $k=4$ for two different samples. This example reveals the only trade-off in the $k$NN algorithm, which is the so-called hyperparameter choice $k$. For small $k$, the assumption of constant density has stronger justification as the spheres are necessarily smaller compared to larger $k$. This, however, comes at the expense of stronger fluctuations, meaning larger statistical uncertainties. These can be systematically reduced by choosing larger $k$, as the volume of the spheres shows a power law in the dimension of the distribution.

Since the estimation of differential entropies from samples is a task of fundamental interest \cite{Hino2015} with applications ranging from statistics \cite{Tarasenko1968, Vasicek1976, Hino2013} to signal processing \cite{Learned-Miller2003, Comon1994} to machine learning and pattern recognition \cite{Mannor2005, Rubinstein2004, Hino2010, Hino2013a}, various other paths beside the aforementioned one have been explored \cite{Czyz2023}. There exists a rich literature for non-parametric entropy estimators that do not assume a specific form of the underlying probability density \cite{Kozachenko1987, Sricharan2013, Singh2016, Singh2003, Kandasamy2015, Joe1989, Hall1993, Gyrfi1987, Goria2005, Gao2018, Birge1995, Berrett2016, Beirlant1997, Lu2020}. These are either based on nearest neighbor statistics or arrive at an estimate for the local density by using kernel density estimates. 
Recently, machine learning-inspired techniques have been explored. Giving up on the non-parametric property of the estimator results in the loss of the asymptotically unbiasedness at possibly increased performance. Numerous works have been put forward in this regard, with applications of estimating entropies \cite{Ao2022} and mutual informations \cite{Belghazi2018}.

\subsection{Benchmarks in low dimensions}
\label{subsec:kNNBenchmarkLowDim}
We showcase the performance of the estimator for Gaussian distributions in \autoref{fig:EntropyEstimatorBenchmark} \textbf{b}). The data points are obtained from 10 independent runs, meaning that a total of $10N_s$ data points were generated, from which we extracted means and standard deviations on the relative deviations $(\hat{S}-S)/S$. We study the performance of the algorithm in $d=2$ dimensions as a function of the squeezing parameter $\gamma$, where $\gamma=0$ corresponds to a normal distribution and large $\gamma$ implies higher squeezing. Here, $S=2.83$ for all values of $\gamma$. A central assumption in the derivation of the expression in Eq.~\eqref{eq:EntropyEstimator} is uniform density inside the $d$-dimensional ball with radius $\epsilon^i(k)$ between sample $i$ and its $k$-th neighbor. For larger squeezing, this assumption is violated more strongly since the squeezing introduces a preferred direction at odds with the assumed isotropy, resulting in worse performance for larger $\gamma$.

\subsection{Scaling with dimension}
\label{subsec:kNNBenchmarkDimScaling}
In \autoref{fig:EntropyEstimatorBenchmark} \textbf{c)}, we benchmark the estimator on a standard normal distribution for increasing dimension. The higher dimension implies that the assumption of constant density inside the $d$-dimensional spheres of radius $\epsilon^i(k)$ must hold for larger volumes if the number of samples is held constant. This will generally result in worse performance, as demonstrated.

\subsection{Tuning away from the Gaussian regime}
\label{subsec:kNNBenchmarkNonGauss}
To test the estimator's performance away from the Gaussian regime, we benchmark it on a generalized normal distribution \cite{Nadarajah2005}. The latter has a probability density of the form
\begin{equation}
    f_{\mathrm{GND}}(x;\mu, \alpha, \beta)=\frac{\beta}{2\alpha\Gamma(1/\beta)}\exp\left(-\left|\frac{x-\mu}{\alpha}\right|^\beta\right),
    \label{eq:GeneralizedNormalPDF}
\end{equation}
with mean $\mu$, scale $\alpha > 0$, shape $\beta > 0$ and $\Gamma$ denoting the gamma function. In \autoref{fig:EntropyEstimatorBenchmark} \textbf{d)}, we vary $\alpha$ and $\beta$ in such a way that the true entropy $S$ is kept constant at the value corresponding to the standard normal distribution in $d=4$ dimensions, for which $\alpha = 1/2, \beta = 2$, i.e. $S=5.67$. We observe the worst performance for small values of $\beta$, i.e., in the regime of heavy tails, which is expected as the central assumption of constant density of the employed estimator needs to hold for larger regions in space. For large values of $\beta$ and sufficiently many samples, the results improve (for $N_s=5000$ (50000) the relative deviations for $\beta=1$, $\beta=3$ and $\beta=9$ are -0.008, -0.007 and 0.003 (-0.004, -0.003 and -0.001)), in line with the intuition that the entropy of a uniform distribution should be easy to estimate. However, for $N_s=50$ and $N_s=500$ samples, this intuition seems to be misleading as the best performance is observed for $\beta=1$ and $\beta=3$, respectively, which resemble Gaussian-like distributions.

\section{Results}
\label{sec:Results}
We are now ready to present our main results: analytical and numerical observations of the area law for classical entropies. After a general comparison of the analytical model and TWA in terms of two-point correlators and entropic quantities, we systematically study typical effects, including non-Gaussian features, a thermal initial state, the influence of the subsystem's position within the total system, the total system size, the dependence of the estimated entropies on the number of samples and finally two types of boundary conditions. 

Unless specified differently, we use the values $c_1=-1/n$, $n=1000$, such that our energy scale is given by $\abs{n c_1} = 1$ (note that this renders \textit{all} parameters and other quantities of our interest dimensionless). Further, we consider Lithium-7 for which $c_0 = - 2 c_1$, set $q=2J$ with $J=2$ and test $N=20$ wells with open boundary conditions at the three different times $t=0.5, 0.75, 1$. To relate these values to SI units, we note that for Lithium-7 at typical denstities $\abs{n c_1}$ is on the order of $100$Hz \cite{Huh2020}, resulting in typical experimental time scales on the order of $10$ms. We note that $t=1$ represents one spin oscillation time, which has been measured to be $t \approx 6$ms \cite{Huh2020}. This is well below the time scale of decoherence effects like three-body loss or atom loss from the trap. From the 20 wells, we study five wells located in the center of the lattice, i.e., wells $8,9,10,11,12$, and neglect the remaining 15 wells from our examination, leaving us with the subsystem depicted in \autoref{fig:SpatialSubregion}. We assume the initial temperature to be zero, i.e., $T=0$, and base the entropy estimation on $10^4$ samples. We systematically vary these parameters one by one and investigate their influences from \autoref{subsec:SystematicsDistributions} to \autoref{subsec:SystematicsBoundaryConditions}. The plotted curves are either dashed or solid, corresponding to interpolations or fits of the (finite-size) area or the volume law.

\begin{figure*}
    \centering
    \includegraphics[width=0.9\linewidth]{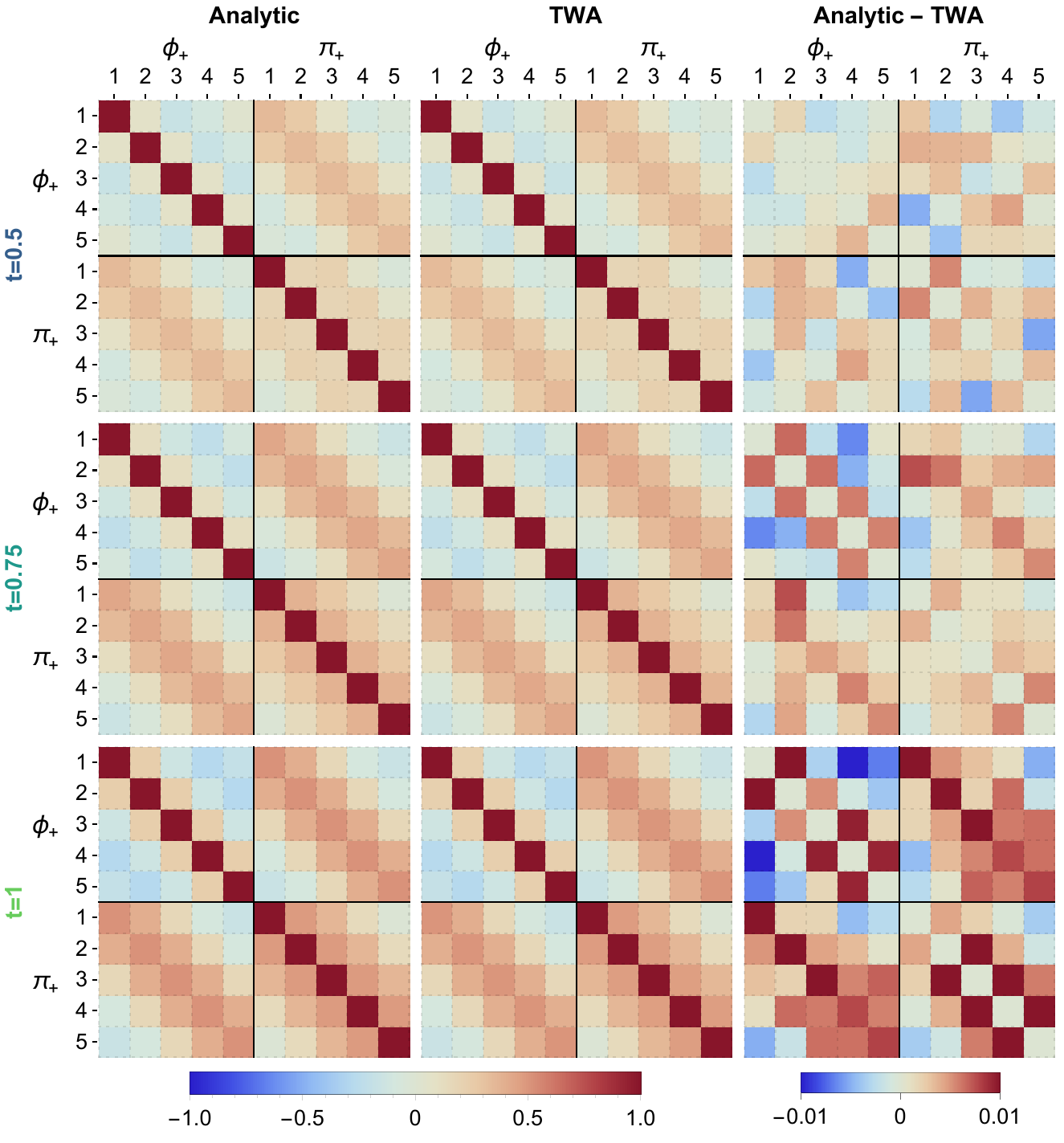}
    \caption{Time evolution of the correlation matrices \eqref{eq:correlationMatrix} generated by the analytic Hamiltonian \eqref{eq:HamiltonianGaussianCanonicalOperators} (left column) and the TWA simulation (middle column) together with their difference (right column).}
    \label{fig:CorrelationMatricesBaseCase}
\end{figure*}

\subsection{Correlation matrices}
\label{subsec:CorrelationMatrices}
As a first qualitative comparison between the analytical model and the truncated Wigner approximation that takes into account all interactions contained in Eq.~\eqref{eq:FullHamiltonian}, we show the correlation matrices between the phase-space variables $\phi_+$ and $\pi_+$ in \autoref{fig:CorrelationMatricesBaseCase} for the five wells under scrutiny at the three different times. The correlation matrix is given by the normalized covariances
\begin{equation}
    \label{eq:correlationMatrix}
    C_+^{j j'} = \frac{\gamma_+^{j j'}}{\sqrt{\gamma_+^{j j} \, \gamma_+^{j' j'}}},
\end{equation}
thereby rendering all entries to lie between -1 and 1.
We want to emphasize that the analytical model is \textit{fully} characterized by the observables in Eq.~\eqref{eq:correlationMatrix}, reducing the complexity of extracting entropies to an extraction of its second moments. In contrast to that, the full Hamiltonian Eq.~\eqref{eq:FullHamiltonian} contains more complex, i.e., higher-order terms, meaning that observables that contain higher-order fluctuations are required to fully characterize entropies.

Since the initial polar state ($t=0$) in \eqref{eq:PolarState} is uncorrelated, its correlation matrix is the identity matrix. During the evolution, correlations between different wells are built up through the transport of atoms permitted by the tunnel Hamiltonian given in Eq.~\eqref{eq:TunnelHamiltonian}. Simultaneously, correlations between  $\phi_+^j$ and $\pi_+^j$ within one well are generated through the internal dynamics in Eq.~\eqref{eq:SingleWellHamiltonian}, leading to the complex structures visible in \autoref{fig:CorrelationMatricesBaseCase}. Remarkably, the analytical model captures most of the features visible in the normalized second-order correlators for the three times considered, with the absolute value of deviations not exceeding $1\%$ (see right column). 

Note, however, that this analysis does not permit any statements about the agreement of higher-order moments. These can, in principle, deviate from a Gaussian prediction due to the presence of fourth-order terms in Eq.~\eqref{eq:FullHamiltonian}. Crucially, we, therefore, do not assume the distribution to obey a specific functional form in our subsequent analysis of sample data generated by the means of TWA with regards to entropic quantities.

The good agreement between the two models leads us to conclude that the analytical model is a justified approximation in regimes where the contribution of fourth-order terms to the full Hamiltonian Eq.~\eqref{eq:FullHamiltonian} is negligible. These regimes are characterized by small values of $c_0$, i.e. Lithium-7, and large values of $q$, such that the undepleted pump approximation (see \autoref{subsec:UndepletedPumpApproximation}) is justified. Furthermore, we expect the approximations made by the analytical model to be valid, especially for early times, which is why we restrict ourselves to times up to $t=1$. An extended discussion for later times can be found in \autoref{subsec:SystematicsDistributions} and in \cite{PRL}.

\begin{figure*}
    \centering
    \includegraphics[width=0.99\linewidth]{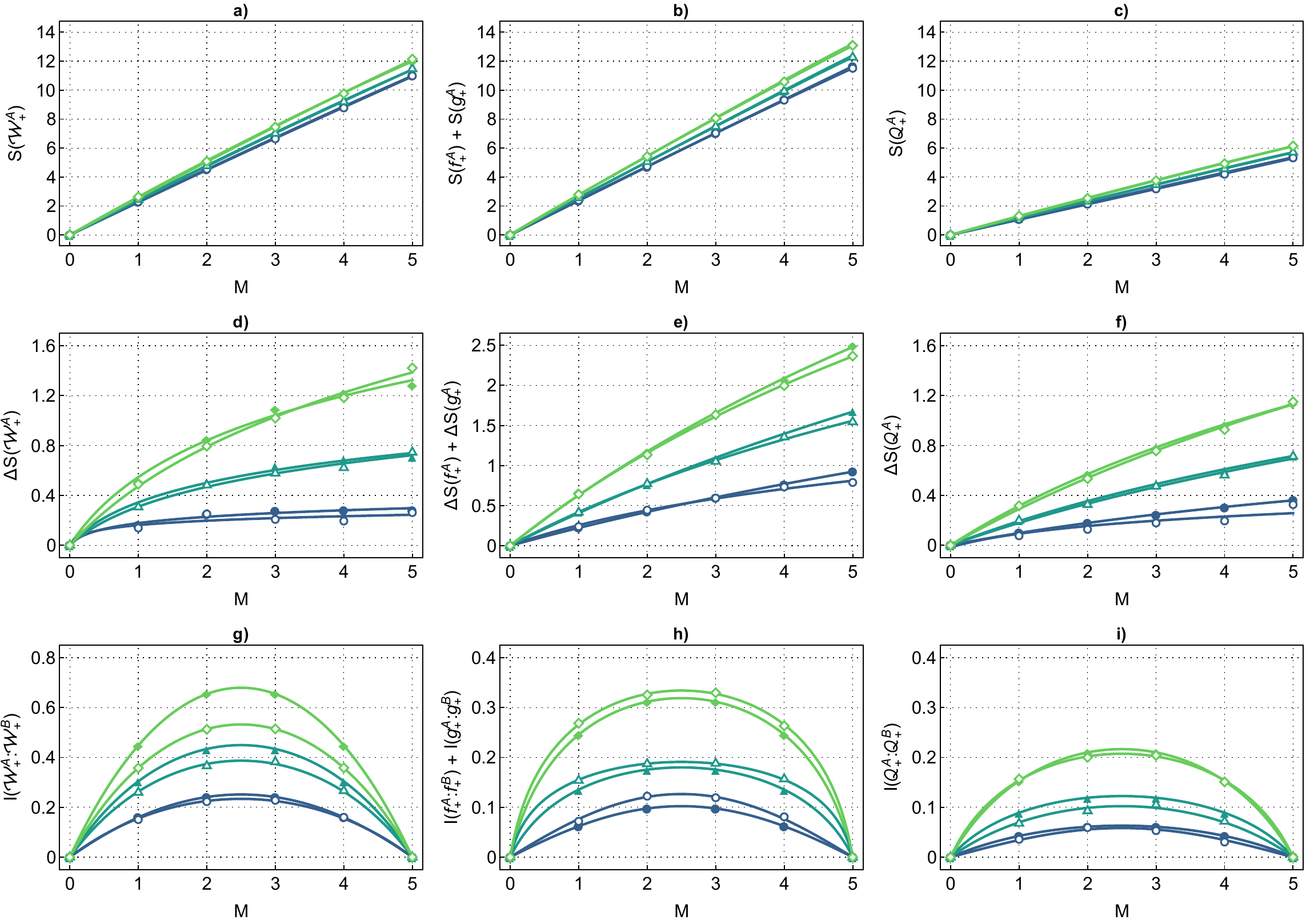}
    \caption{The upper, middle, and lower rows correspond to absolute entropies, subtracted entropies, and mutual information, respectively, while the left, middle, and right columns show Wigner, marginal, and Wehrl quantities. Solid (open) markers correspond to analytical (TWA) data, while solid lines depict fitted curves. The fit function for \textbf{d)} - \textbf{f)} is the area law~\eqref{eq:AreaLaw}, with an additional extensive term for \textbf{a)} - \textbf{c)}. For \textbf{g)} - \textbf{i)}, we fit the finite-size area law \eqref{eq:FiniteSizeAreaLaw}.}
    \label{fig:EntropiesBaseCase}
\end{figure*}

\subsection{Observation of the area law}
\label{subsec:ObservationAreaLaw}
Having gained a qualitative understanding of the dynamics generated by the Hamiltonian \eqref{eq:FullHamiltonian}, we now turn our attention to the dynamics of entropic quantities associated with the quantum state's phase space and measurement distributions. In this context, we want to discuss our main result \autoref{fig:EntropiesBaseCase}, which demonstrates the numerical observation of the area law from experimentally extractable quantities without assuming a specific functional form of the quantum state. Crucially, we only require $10^4$ samples to estimate entropies of up to ten-dimensional distributions, which we deem experimentally feasible. 

In the following, we consider the three distributions of interest, namely 
\begin{itemize}
    \item the Wigner $W$-distribution, introduced in Eq.~\eqref{eq:WignerWDistribution} which  has mainly theoretical interest due to its connection to quantum entropies in the early-time regime and its restricted accessibility,
    \item its marginal distributions, introduced in Eq.~\eqref{eq:MarginalDistributions}, motivated by the fact that they allow for direct experimental extraction as described in \autoref{sec:separate_detection},
    \item and finally the Husimi $Q$-distribution, introduced in Eq.~\eqref{eq:HusimiQDistribution} and directly measurable in experiments using the readout techniques explored in \cite{Kunkel2019,Kunkel2021} and described in \autoref{sec:simultaneous_detection},
\end{itemize}
for subsystem $A$ of varying size $M$, such that $B$ contains $5-M$ wells (see \autoref{fig:SpatialSubregion}). For the aforementioned distributions, we compute the absolute and subtracted entropies as well as the mutual information for the evolution times $t=0.5$, $t=0.75$, and $t=1$ (blue, dark green, light green). We compare the numerical TWA data (open markers) to the analytical data (solid markers) and fit (finite-size) area laws (solid lines).

We first discuss the absolute entropies shown in the first row of \autoref{fig:EntropiesBaseCase}, i.e. \textbf{a)}, \textbf{b)}, \textbf{c)}. All entropies show similar linear behavior, and the area law is masked by the leading order terms. These leading order contributions are subtracted in \autoref{fig:EntropiesBaseCase} \textbf{d)}, \textbf{e)} and \textbf{f)}, as discussed in \autoref{subsubsec:SubtractedClassicalEntropies}, unveiling the area law in form of the typical logarithmic growth in all three types of distributions. Finally, the finite-size area law becomes apparent in all three classical mutual informations, see \autoref{fig:EntropiesBaseCase} \textbf{g)} - \textbf{i)}.

At this point, let us emphasize the challenges associated with the task of estimating the subtracted entropies: We are interested in the estimation of entropies of up to ten-dimensional distributions at an accuracy that is between one and two orders of magnitude higher than the signal itself (consider e.g.  $\Delta S(\mathcal{W}_+^A)/S(\mathcal{W}_+^A) \approx 0.03$ for $M=5$ wells at $t=0.5$). From this perspective, we consider the agreement between analytical and TWA results to be remarkable. 

\begin{figure*}
    \centering
    \includegraphics[width=0.99\linewidth]{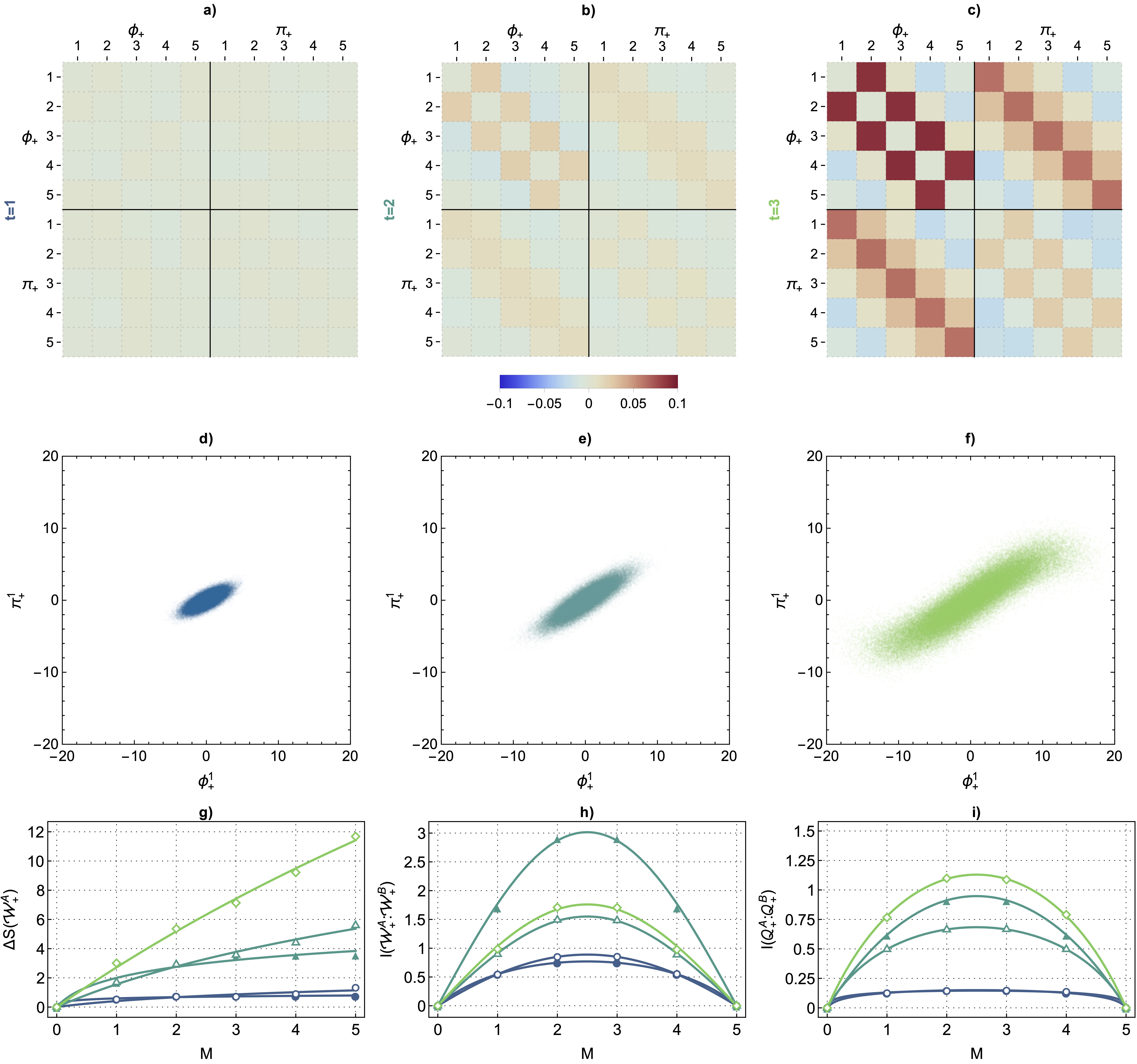}
    \caption{The differences between the analytical model and TWA in the correlation matrices are shown in \textbf{a)}, \textbf{b)} and \textbf{c)} for the three times $t=1$, $t=2$ and $t=3$ (blue, dark green, light green), respectively, and $q=2, J=0.5$. The discrepancies grow as large as 0.1, such that we can no longer meaningfully speak about agreement between the methods. The non-Gaussian form of the underlying phase-space distributions is illustrated in \textbf{d)} - \textbf{f)} in terms of sample points $(\phi_+^1, \pi_+^1)$ of the first well. The corresponding subtracted Wigner entropy, Wigner mutual information, and Wehrl mutual information are shown in \textbf{g)} - \textbf{i)}, for the analytical model (solid markers, only up to $t=2$) as well as TWA (open markers). We observe the area law for the former and the finite-size area law for the two latter quantities, also in the non-Gaussian regime.}
    \label{fig:EntropiesNonGaussian}
\end{figure*}

This concludes the presentation of our main results. In the following, we carry out ablative studies to demonstrate the genericness of our experimental proposal. In particular, we explore later times (\autoref{subsec:SystematicsDistributions}), where the distributions show stronger non-Gaussian features, consider imperfectly prepared, thermal initial states (\autoref{subsec:SystematicsInitialState}) and investigate the influence of boundary effects by shifting the position of the five wells within the total system of $N=20$ wells (\autoref{subsec:SystematicsPosition}). We also reduce the total system to $N=5$ wells to hunt the \textit{finite-size} area law for the subtracted entropies (\autoref{subsec:SystematicsSystemSize}). Finally, we aim to shed light on the sample complexity associated with these tasks when utilizing the $k$NN-estimator (\autoref{subsec:SystematicsSampleSize}) before considering the effect of periodic boundary conditions (\autoref{subsec:SystematicsBoundaryConditions}).

\subsection{Distributions: Gaussian versus non-Gaussian}
\label{subsec:SystematicsDistributions}
To test the influence of non-Gaussian features of the distribution, we consider the parameters $q=2$ and $J=0.5$ at later times, namely $t=2$, $t=3$, and $t=4$. We expect stronger non-Gaussian features to emerge for smaller values of $q$, as we cross the polar to easy-plane ferromagnet phase transition \cite{Stamper2013}, due to the (less detuned) squeezing generated by $c_1$. Similar to \autoref{fig:CorrelationMatricesBaseCase}, we plot the discrepancies between the correlation matrices obtained with the analytical model and TWA for the three different times. \autoref{fig:EntropiesNonGaussian} shows the differences now become more pronounced, growing as large as 0.1, more than an order of magnitude larger compared to \autoref{fig:CorrelationMatricesBaseCase}. We particularly expect the analytical model to mainly give reliable results for the early time dynamics, so a disparity between the two is expected.

To gain more intuition about the characteristics of the distribution at those times, we show samples $(\phi_+^1, \pi_+^1)$ of the Wigner $W$-distribution of the first well in the second row of \autoref{fig:EntropiesNonGaussian}, with all other degrees of freedom integrated out. While the local distribution at $t=2$ still seems relatively Gaussian, an analysis of its 4th-moments reveals that Gaussianity is already strongly violated, with Isserlis' theorem \cite{Cover2006} showing relative deviations as large as 4\%. For $t=3$ ($t=4$), these deviations grow as large as 17\% (60\%). Hence, the entropic quantities also show stronger deviations. In the case of the Wigner $W$-distribution in \autoref{fig:EntropiesNonGaussian} \textbf{g)} substantial differences occur for more wells and late times, as the high dimensionality as well as the strong non-Gaussian features make it difficult for the estimator to reliably estimate the entropies using the $10^4$ available samples. The mutual information shown in \autoref{fig:EntropiesNonGaussian} \textbf{h)} and \textbf{i)} show strong deviations between TWA and the analytical model, although a quantitative comparison does not seem meaningful given the strong non-Gaussian features shown in \autoref{fig:EntropiesNonGaussian} \textbf{d)} - \textbf{f)}. A more informative observation is that the estimates of the mutual informations still change significantly upon increasing the sample size by one order of magnitude, albeit not changing their functional form. In the case of the Wigner mutual information, this underestimation is around 15\% with respect to the currently given data, while it is around 7\% for the Wehrl mutual information. We attribute this difference to the increased smoothness of the Husimi $Q$-distribution compared to the Wigner $W$-distribution.

\begin{figure*}
    \centering
    \includegraphics[width=0.99\linewidth]{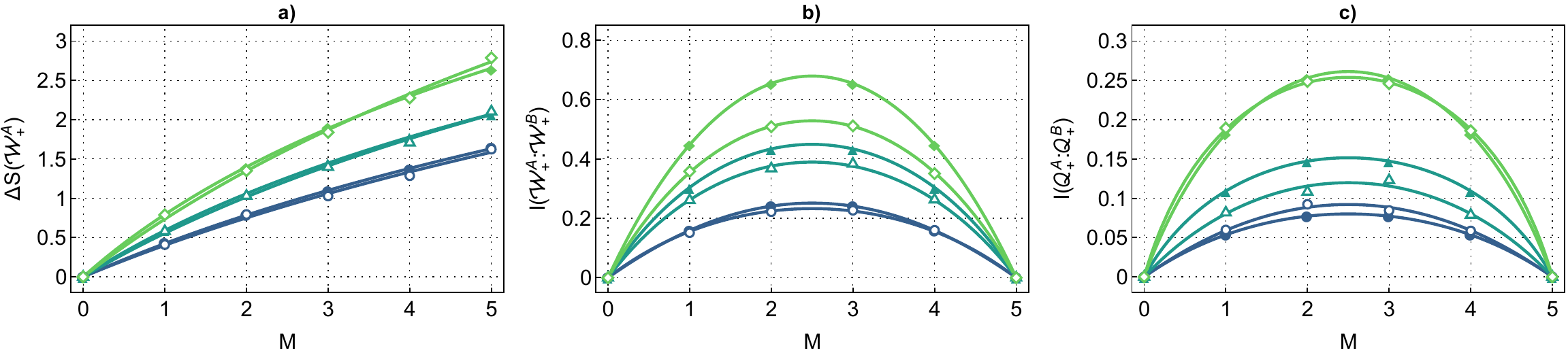}
    \caption{Entropic curves for a thermal initial state with $T=1/2$. The fits correspond to the area law in \textbf{a)} and the finite-size area law in \textbf{b)} and \textbf{c)}, respectively.}
    \label{fig:EntropiesThermal}
\end{figure*}

\begin{figure*}
    \centering
    \includegraphics[width=0.99\linewidth]{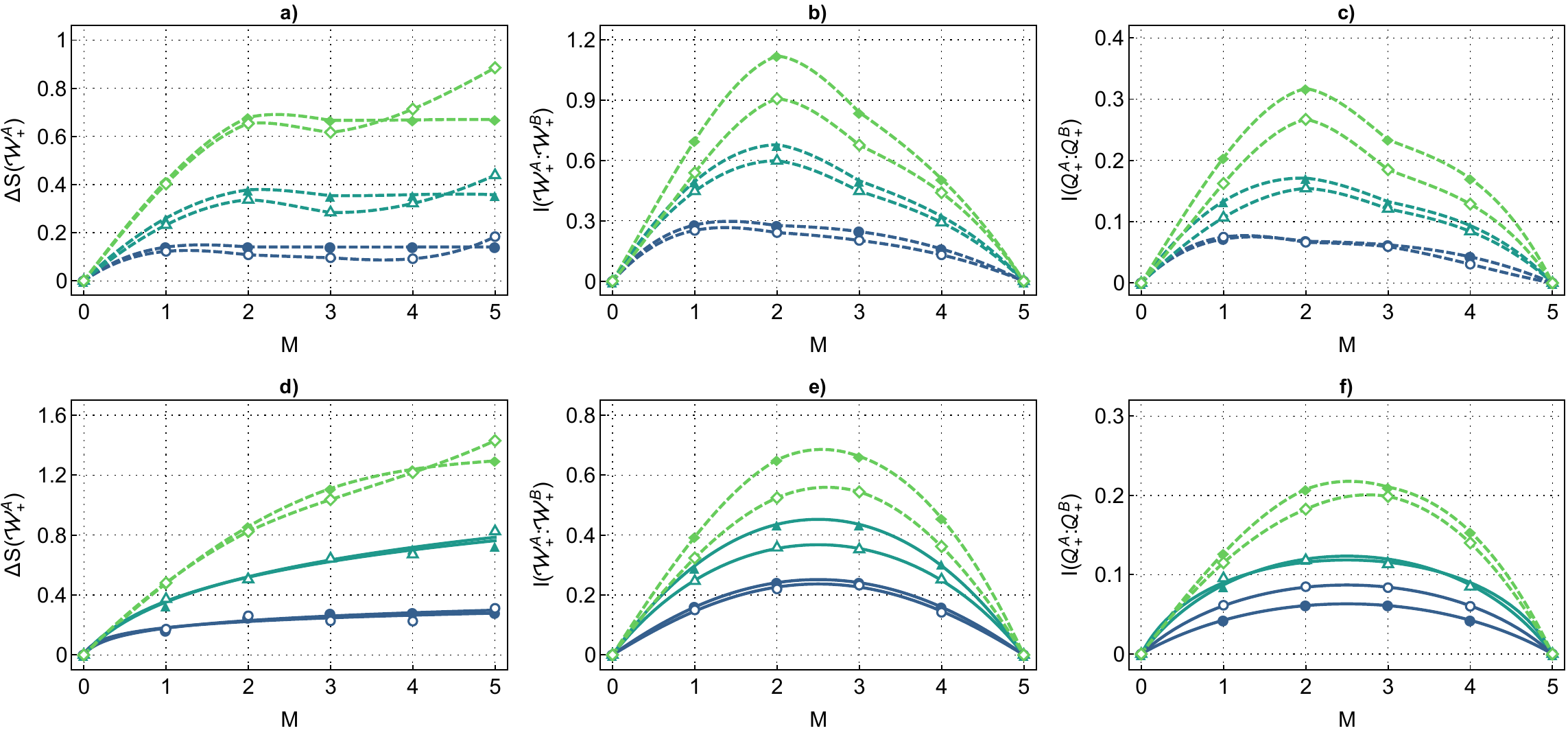}
    \caption{Entropic curves for the wells $M=1-5$ (upper row) and $M=3-7$ (lower row), i.e., when picking five wells out of $N=20$ and partitioning these five wells into two subsystems as shown in \autoref{fig:SpatialSubregion}. Near the boundary, the area laws are strongly distorted by boundary effects, illustrated by the interpolated curves in \textbf{a)} - \textbf{c)}. Further away from the boundary, boundary effects become only relevant when the boundary information had enough time to affect local quantities (note that all curves at $t=1$ are interpolations), see e.g., the asymmetry of the mutual informations in \textbf{e)} and \textbf{f)}.}
    \label{fig:EntropiesPosition}
\end{figure*}

\subsection{Initial state: vacuum versus thermal}
\label{subsec:SystematicsInitialState}
Up to now, our discussion has not been concerned with experimental imperfections such as thermal noise. Here, we want to address this issue by considering thermal initial states that undergo unitary evolution. These initial states are characterized by a temperature scale $\beta$ that we choose to be on the order of a tenth of the system size, such that $\beta = 2$. For a single-well trapping frequency of the order of $100$kHz, this translates to a temperature of $\approx 40$nK. The thermal fluctuations alter the computations of the dynamics by adding thermal noise to the quantum one-half, see Eqs.~\eqref{eq:WignerSamplesVacuumThermal} and \eqref{eq:ThermalAnalytics}.

As expected, the subtracted Wigner entropy $\Delta S(\mathcal{W}_+^A)$ rather features an extensive volume law than an area law, as can be seen in \autoref{fig:EntropiesThermal} \textbf{a)}. When computing both the Wigner and Wehrl mutual information in \autoref{fig:EntropiesThermal} \textbf{b)} and \textbf{c)}, we observe the finite-size area law in both instances, just as in \autoref{fig:EntropiesBaseCase}. The Wigner mutual information shows the same quantitative behavior observed in \autoref{fig:EntropiesBaseCase} \textbf{g)} since the contributions due to the initial thermal noise cancel out. In contrast, the Wehrl mutual information is increased compared to \autoref{fig:EntropiesBaseCase} \textbf{i)}. We attribute this to the inequality \eqref{eq:ClassicalMutualInformationBound} becoming tight in the infinite temperature limit \cite{Lieb2005} (this is also evident from \eqref{eq:CovarianceMatrixHusimiQ}: the additional term $(1/2)\mathds{1}$ becomes irrelevant in this case). Importantly, the finite-size area law \eqref{eq:FiniteSizeAreaLaw} describes the measurable and noisy data well, rendering the proposal robust against thermal noise. Hence, the considered measurement distributions indeed describe a suited setup to observe the area law experimentally. 

\subsection{Subsystem position: center vs. outward}
\label{subsec:SystematicsPosition}
In this subsection, we will choose the position of the subsystem within the bigger system to be wells $M=1-5$ and $M=3-7$ to investigate how the boundary affects the area law.

After computing the subtracted Wigner entropy as well as the Wigner and Wehrl mutual informations for the two subsystem positions, we find the data to be well described by the area law for those instances where boundary effects are irrelevant. The upper row in \autoref{fig:EntropiesPosition} shows that subsystem 1-5 is strongly affected by the boundary on the left, and no area law can be observed. In contrast, the data shown in the lower for the subsystem 3-7 is captured well by the fitted curves, albeit the mutual informations are showing some slight asymmetry for $t=1.0$, i.e., when the boundary information has had sufficient time to propagate to the considered subsystem.

\begin{figure*}
    \centering
    \includegraphics[width=0.99\linewidth]{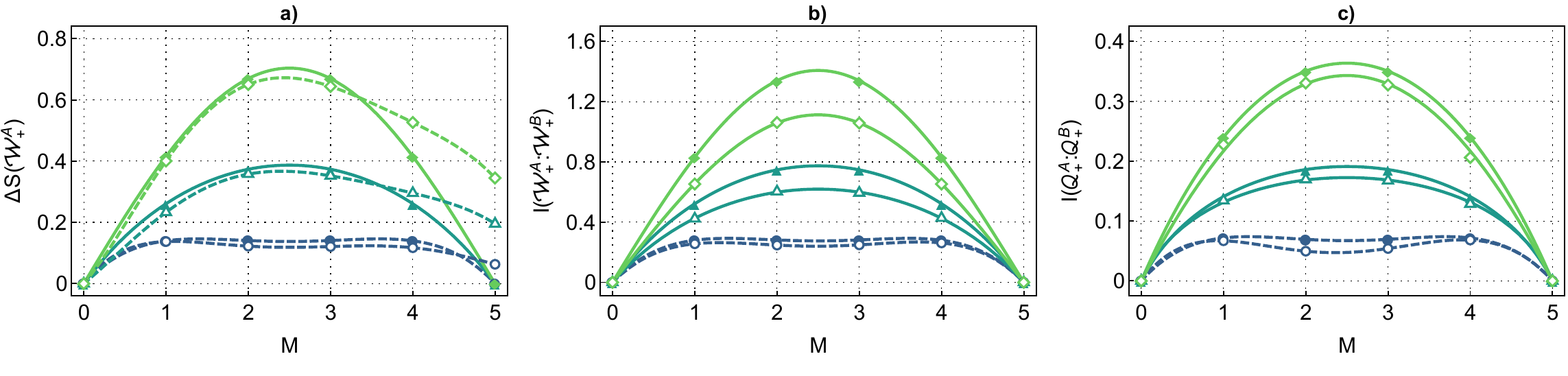}
    \caption{When the total system size equals the subsystem size, i.e., when $N=5$, the area law for the subtracted classical entropies translates into a finite-size area law reflecting the purity of the global state. We remark the underperformance of the entropy estimator for the subtracted Wigner entropies in \textbf{a)} and that the build-up of the finite-size area law only occurs after a transient phase around $t = 0.5$. In both cases, the dashed curves correspond to interpolations, which we show for illustrative purposes.}
    \label{fig:EntropiesSmallSystem}
\end{figure*}

\begin{figure*}
    \centering
    \includegraphics[width=0.99\linewidth]{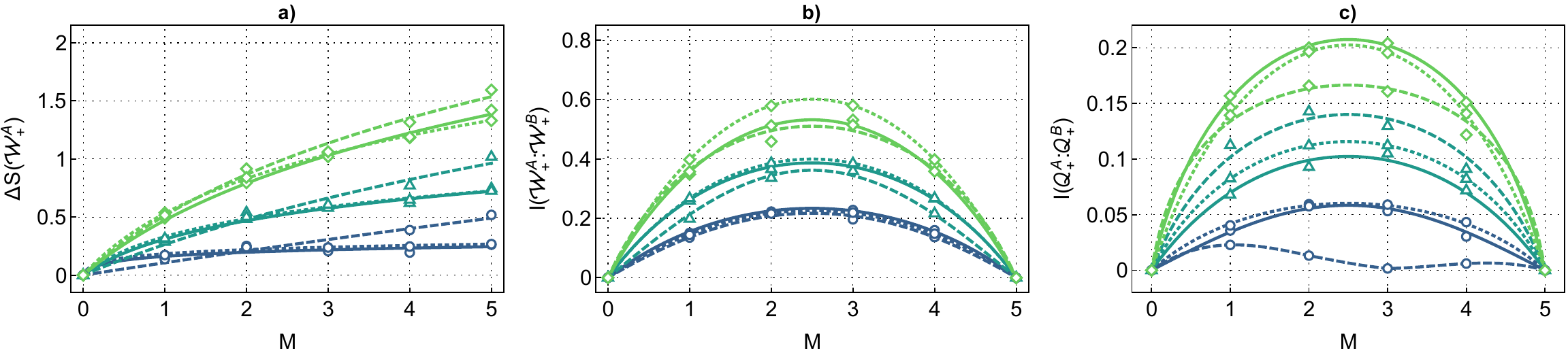}
    \caption{The feasibility of the entropy estimator is checked by varying sample size, where dashed, solid, and dotted lines correspond to $10^3$, $10^4$, and $10^5$ samples. All curves are fits, the only exception being the $10^3$ curve at $t=0.5$ in \textbf{c)}. The convergence between $10^4$ and $10^5$ samples becomes visible especially for the subtracted Wigner entropy in \textbf{a)}  and the Wehrl mutual information in \textbf{c)}.}
    \label{fig:EntropiesSampleSize}
\end{figure*}

\begin{figure*}
    \centering
    \includegraphics[width=0.99\linewidth]{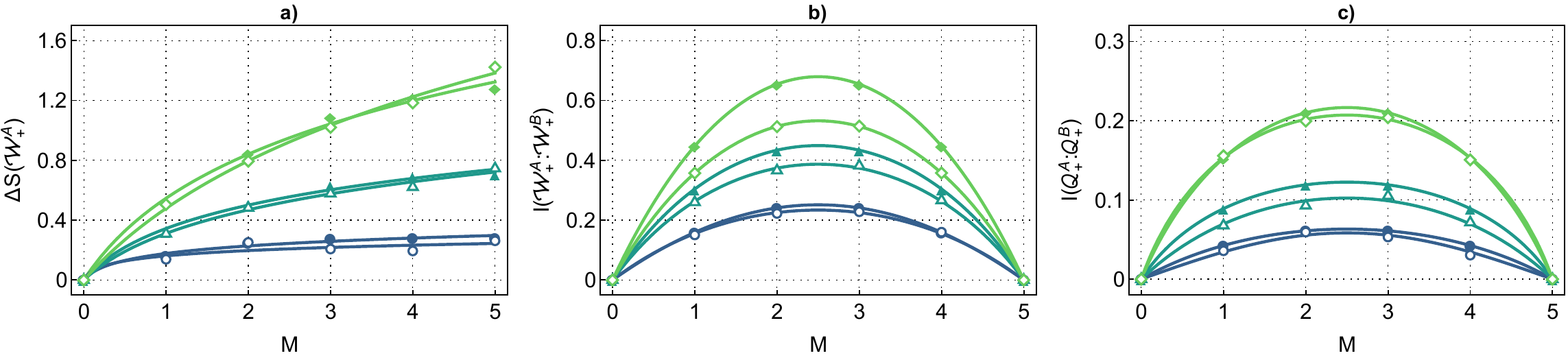}
    \caption{Entropic curves for periodic boundary conditions. As boundary effects are negligible for the subsystem under consideration, all curves agree with those in \autoref{fig:EntropiesBaseCase} to a good extent.}
    \label{fig:EntropiesPBC}
\end{figure*}

\subsection{System size: large vs. small}
\label{subsec:SystematicsSystemSize}

In contrast to the previous discussion, we now restrict the total system to $N=5$ wells such that the evolution within the considered region is unitary. The small system size implies that boundary effects dominate the dynamics, and the regular logarithmic area law is replaced by the finite-size area law, also in the case of the subtracted classical entropies. More precisely, since the density matrix of the five-well system remains pure throughout the evolution, the subtracted Wigner entropy returns to the value zero once the entire system is considered, as shown in \autoref{fig:EntropiesSmallSystem} \textbf{a)}. The reason why this is not captured with the TWA data is due to an insufficient amount of samples. This is in contrast with local distributions, which are likely to be smooth (since local states are likely to be mixed) and hence require fewer samples to be reconstructed adequately. Upon increasing the sample number to $10^5$, the subtracted Wigner entropy is much closer to zero for $M=5$ wells (0.00, 0.06, and 0.18 instead of 0.06, 0.20, and 0.34 for $t=0.50$, $t=0.75$ and $t=1.00$). The Wigner and Wehrl mutual information both feature the area law, but only starting at later times $t \geq 0.75$. 

\subsection{Sample size: small versus large}
\label{subsec:SystematicsSampleSize}
Thus far, we estimated entropies of the distributions generated through TWA using $10^4$ samples without questioning whether the $k$NN estimator had converged for the given sample set. Similar to the benchmarks presented in \autoref{subsec:kNNEstimator}, we want to better understand the sample complexity of the entropy estimation using the $k$NN-estimator. To this end, we carry out the estimations of the subtracted Wigner entropy, the Wigner mutual information and the Wehrl mutual information in \autoref{fig:EntropiesSampleSize} \textbf{a)}, \textbf{b)} and \textbf{c)}, respectively, for $10^3$, $10^4$ and $10^5$ samples, indicated through dashed, solid and dotted lines respectively. 

The subtracted Wigner entropy seems to be converged for all times when using $10^4$ samples, as no significant differences exist to the data points that were obtained using $10^5$ samples. The biggest visual differences exist for the Wigner mutual information for late times, which, however, mainly stem from the different scale between \autoref{fig:EntropiesSampleSize} \textbf{a)} and \textbf{b)}. For the experimentally relevant Wehrl mutual information, we observe convergence for $10^4$ samples for almost all times, with an even further reduced scale compared to \textbf{b)}, rendering a potential experimental implementation feasible.

\subsection{Boundary conditions: open versus periodic}
\label{subsec:SystematicsBoundaryConditions}
While experimental setups will typically use open boundary conditions, a question of theoretical interest is whether the presented framework is sensitive to the type of boundaries that are employed. Therefore, we modify the tunnel Hamiltonian \eqref{eq:TunnelHamiltonian} to also allow atoms to jump from the last to the first well and show the same quantities as previously in \autoref{fig:EntropiesPBC}. Also, in this setting, we find that the area law persists. More precisely, the curves resemble those obtained for open boundary conditions (\autoref{fig:EntropiesBaseCase}) since the considered subsystem is sufficiently far away from the total system's boundaries.

\section{Discussion}
\label{sec:Discussion}
We have demonstrated the experimental feasibility of observing the area law in the subtracted classical entropies and mutual informations of measurable distributions in a multi-well spin-1 Bose-Einstein condensate using numerical techniques. As a testbed, we chose a system consisting of 20 wells that feature internal dynamics as well as tunneling among wells and read out two observables that obey a canonical commutation relation for early times. Focusing on a subregion of five wells, we find (finite-size) area laws in the Wigner distribution, its marginals, as well as the Husimi-Q distribution, where the latter two lend themselves to experimental investigation using independent subsequent measurements \cite{Strobel2014} and a readout technique proposed in \cite{Kunkel2019,Kunkel2021}, respectively. We have shown that all types of subtracted entropies and mutual informations are quantities worth studying in order to observe the area law and have substantiated its robustness with respect to changing various parameters of the experimental setup from \autoref{subsec:SystematicsDistributions} to \autoref{subsec:SystematicsBoundaryConditions}. The discussed area laws are observed dynamically, meaning that the underlying distributions are generated through quenches rather than ground-state preparations, simplifying the experimental requirements. The proposed procedure makes no assumptions regarding the functional form of the distribution under scrutiny by estimating its differential entropy using a suited $k$-nearest-neighbor estimator to locally estimate the density at each sample point. In \autoref{subsec:SystematicsSampleSize}, we demonstrated that for the estimator to converge, we require on the order of $10^4$ samples, which we deem experimentally feasible.

We simulated the system's dynamics using an analytical model derived from the full Hamiltonian and TWA. While the analytical model only features quadratic terms corresponding to an integrable system, the dynamics predicted by the TWA are more intricate and lead to late-time equilibration due to finite particle numbers and the non-integrable nature of the mean-field equations of motion. We accordingly found agreement between the two methods at early times and found pronounced differences at times $t \geq 2$. Particularly for parameter choices that feature stronger non-Gaussian features, such as a lower value of $q$ violating the undepleted pump approximation more strongly, we observed greater disparities between TWA and the analytical model.

In the non-Gaussian regime, we found the $k$NN-estimator to require more samples to reach convergence, hinting at the difficulty of estimating entropies of ten-dimensional distributions while being agnostic to their functional form. As the $k$NN estimator is asymptotically unbiased, it is always possible to increase or decrease the sample size to check whether the estimator has converged. We wish to state that while the $k$NN-estimator has a great appeal due to the guarantee of converging to the true entropy using sufficiently many samples, the estimation of differential entropies is a long-standing challenge, and various ideas have been put forward to tackle the task more efficiently \cite{Kozachenko1987,Gyrfi1987,Joe1989, Hall1993,Birge1995,Beirlant1997,Singh2003,Goria2005,Sricharan2013,Kandasamy2015,Singh2016,Berrett2016,Gao2018,Lu2020,SteegGitHub,Kraskov2004,Hino2015,Tarasenko1968, Vasicek1976, Hino2013,Learned-Miller2003, Comon1994,Mannor2005, Rubinstein2004, Hino2010, Hino2013a,Czyz2023,Ao2022,Belghazi2018}. Of particular interest is the variational approach, which consists of adapting the parameters of a defined function by maximizing the likelihood of the observed samples. While this approach can work well, it is i) generally difficult to build a class of functions that includes the observed density distribution and ii) challenging to converge to the global minimum, rendering the approach uncontrolled. However, future research may aim to build functional forms that are directed at estimating the densities generated by specific classes of Hamiltonians, for which one may hope to arrive at more efficient estimators. Nevertheless, this would come at the expense of losing generality, which may be a significant drawback when applying the methodology to an experiment that suffers from undesired noise effects that one cannot account for analytically.

Having established the presence of quantum features in entropies of classical distributions, one can envision various ways forward. It will be interesting to further investigate which other platforms lend themselves to the herein-discussed approach and obey the area law. Here, systems that are fully described by the set of observables being read out and systems with other types of degrees of freedom that are constrained by algebras different from canonical commutation relations are of interest. For instance, one may consider the hitherto discussed spin-1 BEC in the pseudo-spin 1/2 configuration, in which case three spin operators forming an $\mathfrak{su}(2)$ algebra constitute a complete representation of the system's state with the phase space distribution being represented on a sphere \cite{Zhang1990, Strobel2014}. But also non-atomic setups, such as integrated photonic waveguides, may present promising candidate systems. We emphasize that the approach and analyses presented here are applicable to any setup capable of preparing or dynamically generating an area law.

It will be equally interesting to see which other quantum phenomena can be accessed using classical entropies of measurement distributions. In this context, let us also highlight our recent related work that studies thermalization and the long time limit of the herein discussed system through the lens of classical entropies \cite{PRL} and a work that is concerned with showing the generality of the area law for classical entropies \cite{Haas2023b}. 

\section*{Acknowledgements} 
The authors want to thank Lev Khaykovich for valuable insights on the scattering properties of Lithium-7 and Thomas Gasenzer and Ido Siovitz for helpful discussions during the development of the manuscript. T. H. is supported by the European Union under project ShoQC within the ERA-NET Cofund in Quantum Technologies (QuantERA) program, as well as by the F.R.S.- FNRS under project CHEQS within the Excellence of Science (EOS) program. This work is supported by the Deutsche Forschungsgemeinschaft (DFG, German Research Foundation) under Germany’s Excellence Strategy EXC2181/1-390900948 (the Heidelberg STRUCTURES Excellence Cluster) and within the Collaborative Research Center SFB1225 (ISOQUANT). This work was partially financed by the Baden-Württemberg Stiftung gGmbH. The authors gratefully acknowledge the Gauss Centre for Supercomputing e.V. (www.gauss-centre.eu) for funding this project by providing computing time through the John von Neumann Institute for Computing (NIC) on the GCS Supercomputer JUWELS \cite{JUWELS} at Jülich Supercomputing Centre (JSC).


\appendix

\section{Operator identities for the undepleted pump regime}
\label{app:OperatorIdentitiesUndepletedPump}
We list a few identities to reexpress the Hamiltonian \eqref{eq:TotalHamiltonianUndepletedPump} in terms of the relative modes as well as the canonical variables. For the step from the local modes to the relative modes, we invert \eqref{eq:RelativeModesOperators}, which yields
\begin{equation}
    \boldsymbol{a}_1^j = \frac{1}{\sqrt{2}} \left( \boldsymbol{a}_+^j + \boldsymbol{a}_-^j \right), \quad \boldsymbol{a}_{-1}^j = \frac{1}{\sqrt{2}} \left( \boldsymbol{a}_+^j - \boldsymbol{a}_-^j \right),
\end{equation}
leading to the identities
\begin{widetext}
    \begin{equation}
        \begin{split}
            \boldsymbol{a}_{1}^{j \dagger} \boldsymbol{a}_{1}^{j \dagger} \boldsymbol{a}_{1}^{j} \boldsymbol{a}_{1}^{j} + \boldsymbol{a}_{-1}^{j \dagger} \boldsymbol{a}_{-1}^{j \dagger} \boldsymbol{a}_{-1}^{j} \boldsymbol{a}_{-1}^{j} + 2 \boldsymbol{N}_{1}^{j} \boldsymbol{N}_{-1}^{j} & = \boldsymbol{a}_+^{j \dagger} \boldsymbol{a}_+^{j \dagger} \boldsymbol{a}_+^{j} \boldsymbol{a}_+^{j} + \boldsymbol{a}_-^{j \dagger} \boldsymbol{a}_-^{j \dagger} \boldsymbol{a}_-^{j} \boldsymbol{a}_-^{j} + 2 \boldsymbol{N}_+^{j} \boldsymbol{N}_-^{j}, \\
            \boldsymbol{a}_1^{j \dagger} \boldsymbol{a}_1^{j \dagger} + \boldsymbol{a}_1^{j} \boldsymbol{a}_1^{j} + \boldsymbol{a}_{-1}^{j \dagger} \boldsymbol{a}_{-1}^{j \dagger} + \boldsymbol{a}_{-1}^{j} \boldsymbol{a}_{-1}^{j} &= \boldsymbol{a}_+^{j \dagger} \boldsymbol{a}_+^{j \dagger} + \boldsymbol{a}_+^{j} \boldsymbol{a}_+^{j} + \boldsymbol{a}_{-}^{j \dagger} \boldsymbol{a}_{-}^{j \dagger} + \boldsymbol{a}_{-}^{j} \boldsymbol{a}_{-}^{j}, \\
            \boldsymbol{a}_1^{j} \boldsymbol{a}_{-1}^{j} + \boldsymbol{a}_{1}^{j \dagger} \boldsymbol{a}_{-1}^{j \dagger} &= \frac{1}{2} \left( \boldsymbol{a}^j_{+} \boldsymbol{a}^j_{+} + \boldsymbol{a}^{j \dagger}_{+} \boldsymbol{a}^{j \dagger}_{+} - \boldsymbol{a}^j_{-} \boldsymbol{a}^j_{-} - \boldsymbol{a}^{j \dagger}_{-} \boldsymbol{a}^{j \dagger}_{-} \right), \\
            \boldsymbol{N}_1^{j} - \boldsymbol{N}_{-1}^{j} &= \boldsymbol{a}_+^{j \dagger} \boldsymbol{a}_-^{j} + \boldsymbol{a}_-^{j \dagger} \boldsymbol{a}_+^{j}, \\
            \boldsymbol{N}_1^{j} + \boldsymbol{N}_{-1}^{j} &= \boldsymbol{N}_+^{j} + \boldsymbol{N}_{-}^{j}, \\
            \boldsymbol{a}_{1}^{j \dagger} \boldsymbol{a}_{1}^{j + 1} + \boldsymbol{a}_{1}^{(j +1) \dagger} \boldsymbol{a}_{1}^{j} + \boldsymbol{a}_{-1}^{j \dagger} \boldsymbol{a}_{-1}^{j + 1} + \boldsymbol{a}_{-1}^{(j +1) \dagger} \boldsymbol{a}_{-1}^{j} &= \boldsymbol{a}_{+}^{j \dagger} \boldsymbol{a}_{+}^{j + 1} + \boldsymbol{a}_{+}^{(j +1) \dagger} \boldsymbol{a}_{+}^{j} + \boldsymbol{a}_{-}^{j \dagger} \boldsymbol{a}_{-}^{j + 1} + \boldsymbol{a}_{-}^{(j +1) \dagger} \boldsymbol{a}_{-}^{j}.
        \end{split}
    \end{equation}
\end{widetext}
For the step from relative modes operators to canonical operators, we invert \eqref{eq:CanonicalOperators}, which gives
\begin{equation}
    \boldsymbol{a}^j_{\pm} = \frac{1}{\sqrt{2}} \left( \boldsymbol{\phi}_{\pm}^j + i \boldsymbol{\pi}_{\pm}^j \right),
\end{equation}
implying the identities
\begin{equation}
    \begin{split}
        \boldsymbol{N}^j_\pm &= \frac{1}{2} \left[ \left( \boldsymbol{\phi}_\pm^j \right)^2 + \left( \boldsymbol{\pi}_\pm^j \right)^2 - 1 \right], \\
        \boldsymbol{a}_{\pm}^{j \dagger} \boldsymbol{a}_{\pm}^{j \dagger} + \boldsymbol{a}_{\pm}^{j} \boldsymbol{a}_{\pm}^{j} &= \left( \boldsymbol{\phi}_\pm^j \right)^2 - \left( \boldsymbol{\pi}_\pm^j \right)^2, \\
        \boldsymbol{a}_{\pm}^{j \dagger} \boldsymbol{a}_{\pm}^{j + 1} + \boldsymbol{a}_{\pm}^{(j +1) \dagger} \boldsymbol{a}_{\pm}^{j} &= \boldsymbol{\phi}_\pm^j \boldsymbol{\phi}_\pm^{j+1} +  \boldsymbol{\pi}_\pm^j \boldsymbol{\pi}_\pm^{j+1}.
    \end{split}
\end{equation} 

\hphantom{ShiftBibliographyToNextSide.}
\clearpage


\bibliography{references.bib}

\end{document}